\documentstyle[preprint,pre,eqsecnum,aps,amssymb,epsfig]{revtex}

\begin{document}
\tightenlines
\newcommand{\mS}{{\mathcal S}}
\newcommand{\mV}{{\mathcal V}}
\newcommand{\mL}{{\mathcal L}}
\newcommand{\mM}{{\mathcal M}}
\newcommand{\mK}{{\mathcal K}}
\newcommand{\rv}{{\mathbf r}}

\draft
\title{Wetting-induced effective interaction potential between spherical
  particles} 
\author{C. Bauer, T. Bieker, and S. Dietrich}
\address{Fachbereich Physik, Bergische Universit\"at Wuppertal,\\
D-42097 Wuppertal, Germany}
\date{\today}

\maketitle

\begin{abstract}
Using a density functional based interface displacement model we
determine the effective interaction potential between two spherical
particles which are immersed in a homogeneous fluid such as the
vapor phase of a one-component substance or the A-rich liquid phase of a
binary liquid mixture composed of A and B particles. If this solvent
is thermodynamically close to a first-order fluid-fluid phase
transition, the spheres are covered with wetting films of the incipient
bulk phase, i.e., the liquid phase or the B-rich liquid,
respectively. Below a critical distance between the spheres their
wetting films snap to a bridgelike configuration. We determine phase
diagrams for this morphological transition and analyze its
repercussions on the effective
interaction potential. Our results are accessible to force microscopy
and may be relevant to flocculation in colloidal suspensions.
\end{abstract}


\section{Introduction}
\label{s:intro}

In view of understanding a particular phenomenon in condensed matter,
theory is supposed to identify the corresponding relevant degrees of
freedom and to provide the effective interaction between them by,
approximately, integrating out the remaining ones so that one is left
with a manageable model. It is a major challenge to determine the
effective interactions because that requires to calculate the
partition function of the whole system under the constraint of a fixed
configuration of the relevant degrees of freedom. The benefit for
carrying out this constrained calculation, which in general is more
difficult than the original full problem, is twofold. First, there is
a gain in transparency by describing the system in terms of relevant
degrees of freedom. Secondly, it is typically less risky to apply
approximations for the partial trace because they only concern the
less relevant degrees of freedom.

The determination of the phase behavior and of the structural
properties of multi-component fluids represents a case study for this
general approach. If the composing particles of the mixture are of
comparable size and shape their degrees of freedom have to be treated
on equal footing. The well developed machinery of liquid state
theory~\cite{hansenmcdonald} 
offers various techniques to cope with this problem. However, these
techniques fail to yield reliable results if, e.g., one component is
much larger than the others; in this case numerical simulations become
inefficient and integral theories lose their accuracy. Colloidal
suspensions are a paradigmatic case for such highly asymmetric
solutions. For their description these difficulties can be overcome by
resorting to the general scheme laid out at the beginning with the 
positions of the
colloidal particles as the relevant degrees of freedom. Accordingly
the degrees of freedom of the small solvent particles are to be
integrated out for a fixed configuration of the colloidal particles
which we assume to be smooth, monodisperse spheres. At sufficiently
low concentrations of the suspended particles this leads to an
effective pair potential between them. In many cases the effective
potential resembles the bare one, i.e., the one in the absence of the solvent,
but with modified, effective interaction parameters which depend on
the thermodynamic variables of the system such as pressure and
temperature.

The effective pair potential acquires additional new features if the
solvent is enriched with particles of medium size such as, e.g.,
polymers. If the colloidal particles come close to each other the
depletion zones around them, generated by the finite size of the
medium particles, overlap leading to an entropically driven attraction
of the colloidal particles~\cite{asakuraoosawa,maoetal}. Correlation
effects can modify the form and the range of these depletion forces
considerably~\cite{goetzelmannetal,rothetal}. These effective
potentials have indeed turned out to be successful in describing the
phase behavior of colloidal suspensions~\cite{dijkstraetal}.

Qualitatively new aspects arise if the solvent particles exhibit a
strong cooperative behavior of their own such as a phase transition
which proliferates to the effective potential between the large 
particles. If the solvent undergoes a continuous phase transition,
thermal Casimir forces between the large particles are induced due to
the geometrical constraint they pose for the critical
fluctuations~\cite{eisenriegleretal,hankeetal}. Such forces are
long-ranged and have a strong influence on the phase behavior of the
colloidal particles~\cite{loewen,netz}. If the solvent is
thermodynamically close to a first-order phase transition, wetting
phenomena~\cite{sdreview} can occur at the surfaces of the dissolved
particles (see Ref.~\cite{bieker} and references therein). If the bulk
phase of the solvent is the vapor phase of a one-component fluid, the
surfaces of the large spheres can be covered by a liquidlike wetting
film. This situation corresponds to aerosol particles floating in a
vapor. If the bulk phase of the solvent is the A-rich liquid phase of a
binary liquid mixture composed of (small) A and B molecules, the
dissolved colloidal particles can be coated by the B-rich liquid
phase of the mixture. If the wet spheres approach each other, at a
critical distance the two wetting films snap to a bridgelike
structure. This morphological phase transition is expected to yield a
nonanalytic form of the effective interaction potential between the
large spheres. This nonanalyticity demonstrates that cooperative
phenomena among those degrees of freedom which are integrated out can
leave clearly visible fingerprints on the effective interaction
between the remaining relevant degrees of freedom. The study of this
kind of profileration is not only of theoretical interest in its own
right but seems to play an important (albeit not
exclusive~\cite{lawetal}) role for the experimentally observed
flocculation of colloidal particles dissolved in a binary liquid
mixture close to its demixing transition into an A-rich and a B-rich
liquid
phase~\cite{beysensetc,makeretal,kiralyetal,jayalakshmikaler,gruellwoermann}.
This observation has triggered numerous theoretical efforts devoted to
various possible explanations of it. Since they are reviewed in Sec.~I
of Ref.~\cite{bieker} and more recently in Ref.~\cite{gilipsentejero}
the interested reader is referred to there and 
we refrain from repeating this discussion here.

In our present analysis of this problem we apply density functional
theory~\cite{evans} which offers two advantages. First, this technique
is particularly well suited to calculate, as required here, free
energies under constraints. Secondly, it allows one to keep track of
the basic molecular interaction potentials of the system. We
focus our interest on thermodynamic states of the solvent which are
sufficiently far away from its critical point so that the emerging
liquid-vapor interfaces of the wetting films exhibit only a small
width. Therefore we can apply the so-called sharp-kink approximation
which considers only steplike variations of the solvent density
distribution and thus leaves the interface position as the main
statistical variable. This approximation has turned out to be
surprisingly accurate for the description of wetting
phenomena~\cite{dietnap}. Our analysis extends and goes beyond
previous efforts~\cite{dobbsetal,dobbsyeomans} which are based on a
similar interface displacement model grounded on a phenomenological
ansatz. Whereas Refs.~\cite{dobbsetal} and \cite{dobbsyeomans} are
aimed at mapping out the phase diagram in terms of interaction
parameters for the bridging transition
mentioned above, we focus on the effective interaction potentials
between the wet spheres, which are not presented in
Refs.~\cite{dobbsetal} and \cite{dobbsyeomans}, and on their
microscopic origin. Inter alia, this allows us to compare the
effective interaction potential between the colloidal particles with
the bare one, i.e., in the absence of the solvent, and thus to comment
on the quantitative relevance of the solvent-mediated interaction.
Moreover, we present the phase diagram of the system in terms of the
thermodynamic variables temperature and chemical potential which is
also not contained in Refs.~\cite{dobbsetal} and \cite{dobbsyeomans}.

In Sec.~\ref{s:theory} we describe the implementation of a simple
version of density functional theory for the present problem. For
reasons of simplicity we confine our analysis to liquid-vapor
coexistence of a one-component solvent; the generalization to a binary
solvent is straightforward. In Sec.~\ref{s:morphology} we present some
examples for the numerically calculated wetting film morphologies and
discuss a phase diagram for the aforementioned morphological transition, and
in Sec.~\ref{s:eip} we analyze the effective wetting-induced interaction potential
between the spheres as a function of the distance between the spheres and
the undersaturation. The experimental relevance of our model
calculations is discussed in Sec.~\ref{s:discussion} and
Sec.~\ref{s:summary} summarizes our main results. The Appendix
contains some technical details.

\section{Density functional theory}
\label{s:theory}

\subsection{Model}

\begin{figure} 
\begin{center}
\epsfig{file=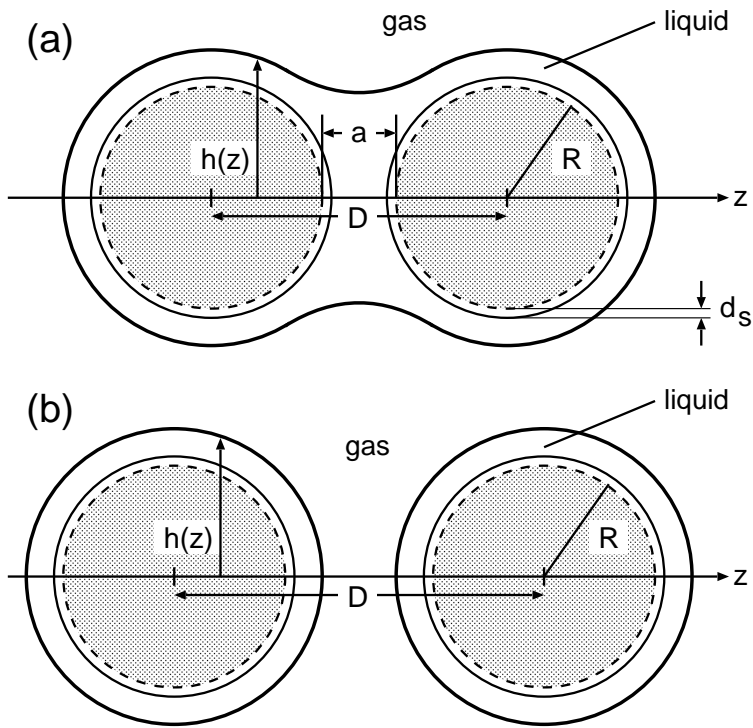, width=10cm}
\end{center}
\caption{\label{f:system}
Wetting film (thick full line) surrounding two identical homogeneous
spheres of radius $R$ which are separated by a distance $D$. The whole
system is rotationally symmetric around the $z$ axis which runs
through both centers of mass. The position of the liquid-vapor interface which
encloses both spheres is described by a function $h(z)$, i.e., in
cylindrical coordinates the sharp interface is given by
the manifold $\{\rv(r_{\perp},\phi,z) =
(r_{\perp}\cos\phi,r_{\perp}\sin\phi,z) \in{\mathbb
  R}^3|r_{\perp}=h(z)\}$. The origin of the coordinate system is in
the middle between the two spheres so that their centers are located
at $z=\pm D/2$. $a=D-2R$ is the shortest separation between the
surfaces of the spheres. Within the so-called sharp-kink
approximation this interface separates a region of
constant liquid number density $\rho_l$ from the surrounding bulk vapor phase
of constant number density $\rho_g$. Close to the surfaces of the
spheres the repulsive 
interaction leads to a volume with thickness $d_s$ excluded for the
centers of the fluid particles. For sufficiently large values of $D$ the bridgelike
wetting film configuration shown in (a) breaks up into two disjunct
pieces so that $h(z) = 0$ for a finite interval around $z=0$ (b).}
\end{figure}

We consider two identical, homogeneous, and smooth spherical particles of radius
$R$ whose centers of mass are separated by a distance $D$ (see
Fig.~\ref{f:system}). They are immersed in a fluid of particles of number
density $\rho(\rv)$ which interact via a Lennard-Jones potential 
\begin{equation}\label{e:ljff}
\phi(r) =
4\epsilon\left(\left(\frac{\sigma}{r}\right)^{12} -
  \left(\frac{\sigma}{r}\right)^6\right).
\end{equation}
The system is symmetric with respect to a rotation around the axis
which connects the centers of mass of the spheres
(Fig.~\ref{f:system}) and with respect to a reflection at a plane in
the middle between the spheres that is perpendicular to the symmetry
axis. Since we work in a grand canonical ensemble and 
the fluid particles are subject to the external potential exerted by
the spheres, the equilibrium number density profile of the fluid
particles exhibits these symmetries, too. Therefore we describe the
system in cylindrical coordinates $(r_{\perp},\phi,z)$, with the $z$ 
axis being the symmetry axis of the system. The 
two centers of mass of the spheres are located at $(r=0,z=\pm D/2)$
such that the spheres occupy the volumes 
$\mS_{\pm} = \{\rv(r_{\perp},\phi,z)
=(x,y,z)=(r_{\perp}\cos\phi,r_{\perp}\sin\phi,z)\in {\mathbb
  R}^3|\pm D/2-R\leq z\leq\pm D/2+R,
\sqrt{r_{\perp}^2+(z\mp D/2)^2}\leq R\}$.
The external potential exerted by both spheres on each individual fluid
particle is 
\begin{equation}\label{e:vtot}
v_{tot}(r_{\perp},z;R) = v\left(\sqrt{r_{\perp}^2+(z-D/2)^2};R\right) +
v\left(\sqrt{r_{\perp}^2+(z+D/2)^2};R\right) 
\end{equation}
where (see Eq.~(A.4) in Ref.~\cite{bieker})
\begin{eqnarray}\label{e:vsphere}
v(r;R) & = &
\frac{9}{8}u_9\left(\frac{1}{r(r+R)^8}-\frac{1}{r(r-R)^8}\right)
-u_9\left(\frac{1}{(r+R)^9}-\frac{1}{(r-R)^9}\right) \nonumber\\
& & -\frac{3}{2}u_3\left(\frac{1}{r(r+R)^2}-\frac{1}{r(r-R)^2}\right)
+u_3\left(\frac{1}{(r+R)^3}-\frac{1}{(r-R)^3}\right)
\end{eqnarray}
is the interaction potential between a single sphere of radius $R$ and a fluid
particle at a distance $r>R$ from the center of mass of the
sphere. In a continuum description, $v(r;R)$ follows from
an integration of the Lennard-Jones potential
\begin{equation}\label{e:ljsphere}
\phi_{sf}(r) = 4\epsilon_{sf}\left(\left(\frac{\sigma_{sf}}{r}\right)^{12} -
  \left(\frac{\sigma_{sf}}{r}\right)^6\right)
\end{equation}
between a molecule of the spherical \emph{s}ubstrate and a \emph{f}luid particle.
The subscript $sf$ denotes the parameters
of the dispersion interaction between a particle in the fluid and a
particle in the spheres. One has $u_3 =
\frac{2\pi}{3}\epsilon_{sf}\rho_s\sigma_{sf}^6$ and $u_9 =
\frac{4\pi}{45}\epsilon_{sf}\rho_s\sigma_{sf}^{12}$ where $\rho_s$ is
the number density of the particles forming the spheres. (Many colloidal
particles exhibit an even more complicated substrate potential because
they are coated by a material different from their core so that they
are no longer radially homogeneous as assumed for Eq.~(\ref{e:vsphere}).)

Within our density functional approach the equilibrium particle
number density distribution of 
the inhomogeneous fluid surrounding the spheres in a grand canonical
ensemble minimizes the functional~\cite{evans}
\begin{eqnarray}\label{e:ldafunc}
\Omega([\rho(\rv)];T,\mu) & = & \int\limits_{\mV_f} d^3r
\Big(f_{HS}(\rho(\rv),T) + \big(v_{tot}(\rv)-\mu\big)\rho(\rv)\Big) \nonumber\\
& & + \frac{1}{2}\int\limits_{\mV_f}\int\limits_{\mV_f} d^3r\,d^3r'\,
w(|\rv-\rv'|)\rho(\rv)\rho(\rv'). 
\end{eqnarray}
$\mV_f = \mV\setminus(\mS_+\cup\mS_-)$ is the
volume accessible for the fluid particles, $\mV$ is the total volume of
the system; $\mV \to {\mathbb R}^3$ in the thermodynamic
limit. Equation~(\ref{e:ldafunc}) does not include the bare
interaction potential $\Phi(D;R)$ (see, c.f., Sec.~\ref{s:discussion})
between the solid spheres, separated by vacuum, generated by the
dispersion forces between the molecules forming the two spheres.
$f_{HS}(\rho,T)$ is the free energy density
of a hard-sphere fluid of number density $\rho$ at temperature $T$. In
Eq.~(\ref{e:ldafunc}), the hard-sphere reference fluid is treated in
local density approximation. We apply the
Weeks-Chandler-Andersen procedure \cite{wca} to split up  
$\phi(r)$ into an attractive part $\phi_{att}(r)$ and a repulsive
part $\phi_{rep}(r)$. The latter gives rise to an effective,
temperature dependent hard-sphere diameter
\begin{equation}
d(T) = \int\limits_0^{2^{1/6}\sigma}dr\,
\left(1-\exp\left(-\frac{\phi_{rep}(r)}{k_BT}\right)\right) 
\end{equation}
which is inserted into the Carnahan-Starling expression~\cite{cs}
\begin{equation}\label{e:cs}
f_{HS}(\rho,T) =
k_BT\rho\left(\ln(\rho\lambda^3)-1+\frac{4\eta-3\eta^2}{(1-\eta)^2}\right)
\end{equation}
for the free energy density $f_{HS}$ of the \emph{h}ard-\emph{s}phere fluid, where
$\eta=\frac{\pi}{6}\rho(d(T))^3$ is the dimensionless packing 
fraction and $\lambda$ is the thermal de Broglie wavelength.
We approximate the attractive part of the interaction $\phi_{att}(r)$ by
\begin{equation}\label{e:attractivepot}
w(r) = \frac{4w_0\sigma^3}{\pi^2}(r^2+\sigma^2)^{-3}
\end{equation}
with
\begin{equation}\label{e:w0}
w_0 = \int_{{\mathbb R}^3}d^3r\,w(r) = \int_{{\mathbb
R}^3}d^3r\,\phi_{att}(r) = -\frac{32}{9}\sqrt{2}\pi\epsilon\sigma^3
\end{equation}
in order to simplify subsequent analytical calculations. The double
integral in Eq.~(\ref{e:ldafunc}) takes into account this attractive
interaction within mean-field approximation.

In the bulk the particle density $\rho_{\gamma}$ (where
$\gamma=l,g$ denotes the liquid and vapor phase, respectively) is
spatially constant, leading to (see Eq.~(\ref{e:ldafunc}))
\begin{equation}\label{e:bfedensity}
\Omega_b(\rho_{\gamma},T,\mu) = f_{HS}(\rho_{\gamma},T) +
\frac{1}{2}w_0\rho_{\gamma}^2 - \mu\rho_{\gamma}
\end{equation}
for the grand canonical free energy density of the \emph{b}ulk
fluid. Minimization of 
$\Omega_b$ with respect to $\rho_{\gamma}$ yields the equilibrium densities.
The line $\mu=\mu_0(T)$ of bulk liquid-vapor coexistence and the two
bulk densities $\rho_l$ and $\rho_g$ at coexistence follow from
\begin{equation}\label{e:coexcond}
\left.\frac{\partial\Omega_b}{\partial\rho}\right|_{\rho=\rho_g} = 
\left.\frac{\partial\Omega_b}{\partial\rho}\right|_{\rho=\rho_l} = 0
\quad \mbox{and} \quad \Omega_b(\rho_g) = \Omega_b(\rho_l).
\end{equation}
For $\mu\neq\mu_0$, i.e., off coexistence, only the liquid or the
vapor phase is stable. In this case the density of the metastable
phase corresponds to the second local minimum of $\Omega_b$.

\subsection{General expressions for the contributions to the effective
  interaction potential}

Henceforth we consider the case that the substrate is sufficiently
attractive so that the liquid phase
is preferentially adsorbed. Therefore, if in the bulk the vapor phase is stable
($\mu\leq\mu_0$), the fluid density is significantly increased in
the vicinity of both spheres. In the spirit of the so-called
sharp-kink approximation (see Sec.~\ref{s:intro} and
Ref.~\cite{dietnap}) we assume that a thin film of constant density 
$\rho_l$ but with locally varying thickness is adsorbed at the
surfaces of the spheres, separating the 
spheres from the bulk vapor phase of density $\rho_g$. This wetting film
encapsulating both spheres is characterized by a function $h(z)$:
\begin{equation}\label{e:hz}
\rho(\rv) = \rho(r_{\perp},\phi,z) = 
\Theta(r_{\perp}-(R+d_s))\bigg(\Theta(h(z)-r_{\perp})\rho_l +
\Theta(r_{\perp}-h(z))\rho_g\bigg) 
\end{equation}
where $\Theta$ denotes the Heaviside step function. The length $d_s$
takes into account the excluded volume at the surfaces of the spheres
which the centers of the fluid particles cannot penetrate due to repulsive
forces. The profile $h(z)$ as given by
Eq.~(\ref{e:hz}) can describe both a 
configuration in which the wetting films surrounding each sphere are
connected by a liquid bridge as well as the configuration in which
both single spheres are surrounded by disjunct wetting layers. In the
latter configuration there is a region around $z=0$ with
$h(z)=0$.

Inserting $\rho(r_{\perp},\phi,z)$ from Eq.~(\ref{e:hz}) into the
functional $\Omega$ in Eq.~(\ref{e:ldafunc}) leads to a decomposition of
$\Omega = \mbox{Vol}(\mV_f)\Omega_b(\rho_g)+\Omega_S$ into a bulk
and subdominant contributions. The bulk 
contribution is $\mbox{Vol}(\mV_f)\Omega_b(\rho_g)$ (with
$\Omega_b$ given by Eq.~(\ref{e:bfedensity})) and
corresponds to the vapor phase which is stable in the bulk. The subdominant
contribution is 
\begin{equation}\label{e:decomp}
\Omega_S[h] = \Omega_{sl} + \Omega_{ex}[h] + \Omega_{ei}[h] +
\Omega_{lg}[h]
\end{equation}
where only $\Omega_{sl}$ is independent of $h(z)$ and all the other three
contributions are functionals of $h(z)$. Since we have not found an
indication for spontaneous symmetry breaking, in the following we
discuss only symmetric configurations with $h(z) = h(-z)$.
\begin{equation}\label{e:excess}
\Omega_{ex}[h(z)] = \mbox{Vol}(\mL)\Big(\Omega_b(\rho_l)-\Omega_b(\rho_g)\Big) 
\end{equation}
with 
\begin{equation}\label{e:volliq}
\mbox{Vol}(\mL) = 2\pi\int\limits_0^{L_z}dz\,h^2(z) -
\frac{8\pi}{3}R^3
\end{equation}
is an \emph{ex}cess contribution which takes into account that the volume
$\mL = \mK\setminus(\mS_-\cup \mS_+)$ is filled with the metastable
liquid instead of the vapor phase; $\mK = \{ \rv(r_{\perp},\phi,z) \in {\mathbb
  R}^3 | r_{\perp}\leq h(z)\}$ 
is the volume enclosed by the liquid-vapor interface. (The excluded
volume due to $d_s$ enters into $\Omega_{sl}$ (see, c.f.,
Eq.~(\ref{e:omegasl})).) This free energy
contribution $\Omega_{ex}$ vanishes at two-phase coexistence
$\mu=\mu_0(T)$ (compare Eq.~(\ref{e:coexcond})).
$2L_z$ is the extension of the total volume of the system $\mV$ in $z$ direction;
$L_z\to\infty$ in the thermodynamic limit and $h(z>z_{max}) = 0$ with
$z_{max} \ll L_z$.
\begin{equation}\label{e:omegaei}
\Omega_{ei}[h(z)] = 2\Delta\rho\int\limits_{\mV_-\setminus \mK_-}
d^3r\,\Big(\rho_l(t(\rv,\mS_-)+t(\rv,\mS_+)) -
v_{tot}(\rv)\Big) 
\end{equation}
can be interpreted as the integrated \emph{e}ffective \emph{i}nteraction between the
spheres and the liquid-vapor interface described by $h(z)$;
$\Delta\rho = \rho_l-\rho_g$. $\mV_-$ is that part of the volume
$\mV$ with $z<0$ (we note again that $\mV_-\to{\mathbb R}_-^3$ in the
thermodynamic limit which is always considered here), 
analogously $\mK_-$ is the part of the set $\mK$ with $z<0$. In
Eq.~(\ref{e:omegaei}) we have introduced the interaction potential
\begin{equation}
t(\rv;\mM) = \int\limits_{\mM} d^3r'\,w(|\rv-\rv'|)
\end{equation}
between a fluid particle at $\rv$ and a region $\mM$ (with
$\rv\not\in\mM$) homogeneously filled with the same fluid
particles (analogous to the function $t(z)$ introduced in
Refs.~\cite{sdreview} and \cite{dietnap} in the case of a planar
substrate). $v_{tot}$ is the total interaction potential between a 
fluid particle and both spheres (see Eq.~(\ref{e:vtot})).
Finally,
\begin{equation}\label{e:omegalg}
\Omega_{lg}[h(z)] = -(\Delta\rho)^2\int\limits_{\mV_-\setminus
  \mK_-} d^3r\,\Big(t(\rv;\mK_-)+t(\rv;\mK_+)\Big)
\end{equation}
is the free energy contribution from the free \emph{l}iquid-\emph{g}as interface. It
is a \emph{nonlocal} functional of $h(z)$ in contrast to $\Omega_{ex}$
and $\Omega_{ei}$ whose dependence on $h(z)$ enters only via the
integration volume $\mK_-$. The \emph{local} approximation thereof,
which is provided by the gradient expansion of
Eq.~(\ref{e:omegalg}), is
\begin{equation}\label{e:local}
\Omega_{lg}^{loc} = 4\pi\sigma_{lg}^{(p)}\int\limits_0^{L_z} dz\, h(z)\,
\sqrt{1+\left(\frac{dh}{dz}\right)^2}.
\end{equation}
In Eq.~(\ref{e:local})
\begin{equation}\label{e:sigmalgplanar}
\sigma_{lg}^{(p)} =
-\frac{1}{2}(\Delta\rho)^2\int\limits_0^{\infty}dz\,
\int\limits_z^{\infty}dz'\int\limits_{{\mathbb R}^2}d^2r_{\parallel}
w\left(\sqrt{r_{\parallel}^2+z'^2}\right)
\end{equation}
is the interfacial tension of a planar, free liquid-vapor interface in
sharp-kink approximation. We note that, strictly speaking, the
surface tension of a curved liquid-vapor interface depends on the local
radius of curvature (see Fig.~2 in Ref.~\cite{bieker} and the
references therein concerning the Tolman length). This curvature
dependence is omitted in the local model presented here. However, for
spheres of radius $R\geq20\sigma$ as considered henceforth the curvature 
correction is less than $1\%$. Similar arguments hold for the
deviation of the actual liquidlike density in the wetting film from
the bulk value $\rho_l$.

For our choice of interaction potentials (Eq.~(\ref{e:ljff}) and 
(\ref{e:ljsphere})) a tedious calculation leads to explicit
expressions for the contributions $\Omega_{ei}$ and
$\Omega_{lg}$ which are given in the Appendix. The remaining contribution 
\begin{equation}\label{e:omegasl}
\Omega_{sl} = -\rho_l\int\limits_{\mV_-\setminus \mS_-}
d^3r\,\Big(\rho_l(t(\rv,\mS_-)+t(\rv,\mS_+)) -
2v_{tot}(\rv)\Big) - \Omega_b(\rho_l)\frac{8\pi}{3}((R+d_s)^3-R^3),
\end{equation}
which is independent of $h(z)$, is the \emph{s}phere-\emph{l}iquid interfacial free
energy corresponding to the interface between the spheres and the
liquid phase. The last term in Eq.~(\ref{e:omegasl}) takes into
account the excluded volumes at the surfaces of the spheres. In the
limit of large separations $D$ one has
\begin{equation}\label{e:omegasl_asymp}
\Omega_{sl}(D\to\infty) - 2\Omega_{sl}^{(1)} \sim D^{-6}
\end{equation}
with the sphere-liquid interfacial free energy $\Omega_{sl}^{(1)}$ of
a single sphere immersed in the liquid phase. The
leading power law $\sim D^{-6}$ in Eq.~(\ref{e:omegasl_asymp}) can be
inferred from the following consideration: 
if present, the second sphere displaces a spherical volume from the
homogeneous liquid phase so that the free energy of the interaction of
the first sphere with the bulk liquid is reduced by the interaction
free energy of that sphere with the displaced spherical liquid volume. This
latter interaction decays as $D^{-6}$ for large separations $D$, at
which the dispersion interaction between two spherical objects resembles the
dispersion interaction between two pointlike particles. (Here, as
before, we have not yet taken into account the bare interaction
potential $\Phi(D;R)$ between the two solid spheres; but see, c.f.,
Sec.~\ref{s:discussion}.) 

Up to the bulk contribution the grand canonical potential of the
system is the minimum of $\Omega_S[h(z)]$ with respect to the profile $h(z)$:
\begin{equation}\label{e:gcpismin}
\Omega_S = \Omega_S(D;R) = \min_{\{h(z)\}}(\Omega[h(z)]).
\end{equation}
Thus the equilibrium interface morphology $h(z)$ minimizes
$\Omega_S[h(z)]$ which includes the contributions $\Omega_{ex}[h(z)]$,
$\Omega_{ei}[h(z)]$, $\Omega_{lg}[h(z)]$, and $\Omega_{sl}$.
The functional used in Refs.~\cite{dobbsetal} and
\cite{dobbsyeomans} (Eq.~(1) in both references) is, albeit formulated
in another coordinate system and using a more phenomenological ansatz
for the basic interaction potentials, essentially identical with the sum
$(\Omega_{lg}^{loc}+\Omega_{ex}+\Omega_{ei})[h(z)]$. However, this model 
description does neither incorporate the bare dispersion
interaction of the two spheres (c.f., Sec.~\ref{s:discussion}) nor the
free energy contribution 
$\Omega_{sl}$ which describes the sphere-liquid interfacial
free energy. We emphasize that the consideration of the contribution
$\Omega_{sl}$ -- which does not depend on $h(z)$ -- is not essential
for the determination of the equilibrium wetting film morphology and
hence it is not relevant for the thermodynamic phase diagram of
thin-thick and bridging 
transitions (Fig.~2 in Ref.~\cite{dobbsyeomans}) for a \emph{fixed}
separation $D$ between the spheres. But the term $\Omega_{sl}$ is crucial to
the \emph{shape} of the effective,
wetting-induced interaction potential between the spheres, i.e., its
dependence on $D$ (see Eq.~(\ref{e:omegasl_asymp})).

\section{Morphology of the wetting layers}
\label{s:morphology}

\subsection{Interface profiles}

The actual wetting layer morphology $h(z)$ follows from numerical
minimization of the functional $\Omega_S[h(z)]$ (Eq.~(\ref{e:decomp}))
for a given temperature $T$ and undersaturation $\Delta\mu=\mu_0(T)-\mu$,
with the contributions $\Omega_{ex}$ (Eq.~(\ref{e:excess})),
$\Omega_{ei}$ (Eq.~(\ref{e:omegaei})), $\Omega_{sl}$
(Eq.~(\ref{e:omegasl})), and $\Omega_{lg}$ (Eq.~(\ref{e:omegalg}) within
the nonlocal and Eq.~(\ref{e:local}) for the local theory). Within
a range of parameters $(T,\Delta\mu)$ the numerical minimization
yields two different solutions for $h(z)$, one with a liquid bridge and one
without, depending on the initial function $h(z)$ used in the
iteration scheme for the minimization. For small separations
$a\ll2R$ only the solution which exhibits
a liquid bridge is stable whereas for large separations
$a\gg2R$ only the solution without bridge minimizes $\Omega_S$. For
large distances $D$ the minimization consistently
yields twice the result known for a single individual sphere enclosed by a wetting
film (compare Ref.~\cite{bieker}). This observation amounts to a
useful check of the numerical procedure.

\begin{figure} 
\begin{center}
\epsfig{file=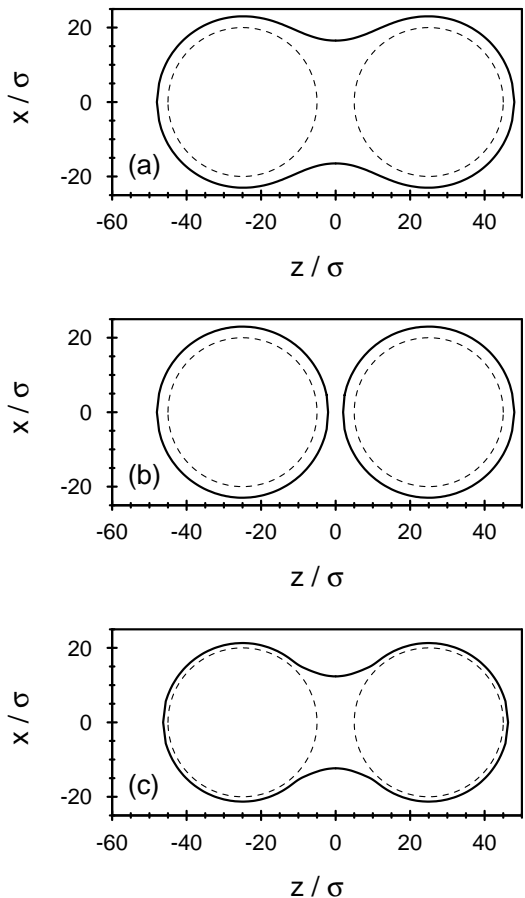, width=7cm, bbllx = 15, bblly = 15,
  bburx = 495, bbury = 840}
\end{center}
\caption{\label{f:example1}
Morphologies of liquidlike wetting layers on two adjacent, identical
spheres with radius $R=20\sigma$. The center-of-mass distance between
them is $D=50\sigma$. The pictures show cross-sections through the system
defined by the plane $y=0$; the system is rotationally symmetric
around the $z$ axis (see Fig.~\ref{f:system}). The thick full lines denote the
liquid-vapor interface, the 
thin dashed lines the surfaces of the spheres. (a) and (b): layer
configuration with and without liquid bridge, respectively, for the
temperature $T^*=k_BT/\epsilon = 1.3>T_{tt}^*$ and at liquid-vapor
coexistence $\Delta\mu=0$. Because of its higher free energy the configuration
without bridge is metastable (c.f., Fig.~\ref{f:btandttt}). These
configurations are characterized by the interaction potential parameters 
$u_3 = 6.283\epsilon\sigma^3$, $u_9 =
0.838\epsilon\sigma^9$, and $d_s=\sigma$. The temperature is above the thin-thick
transition temperature $T_{tt}^*\approx1.271$ for each single sphere. In (c)
the interaction 
parameters and $D$ are the same, but the temperature
$T^*=1.2$ is below the thin-thick transition temperature $T_{tt}^*$, so that the
wetting layer around a single sphere is thinner than in (a) and
(b). Also at this temperature the bridge configuration is the stable
one (c.f., Fig.~\ref{f:btandttt}).} 
\end{figure}

As a first example, in Fig.~\ref{f:example1} we present the numerical
results for a wetting layer enclosing two spheres of radius
$R=20\sigma$. For our particular choice of interaction parameters, at
coexistence $\Delta\mu=0$ the wetting film on each of the single spheres alone
exhibits a first-order \emph{t}hin-\emph{t}hick transition (which is the remnant
of the first-order wetting transition on the corresponding planar
substrate, see Fig.~8(a) in Ref.~\cite{bieker}) at $T_{tt}^* =
k_BT/\epsilon \approx 1.271$ 
(which corresponds to $T_{tt}/T_c \approx 0.9$ where $T_c$ is the
critical temperature of gas-liquid coexistence in the bulk). The planar substrate,
i.e., a single sphere in the limit $R\to\infty$, exhibits a genuine
first-order wetting transition (with the film thickness jumping to
a macroscopic value) at $T_w^* \approx 1.053$ ($T_w/T_c \approx 0.75$,
$T_{tt}/T_w\approx1.21$). Figure~\ref{f:example1}(a) depicts a typical
solution with a bridge, here for a separation $a = D-2R = 10\sigma$ ($D = 50\sigma$)
and the thermodynamic parameters $T^*=1.3>T_{tt}^*$ and
$\Delta\mu=0$, i.e., at liquid-vapor coexistence.
The solution without a bridge for the same
choice of parameters is shown in Fig.~\ref{f:example1}(b). The latter
solution has a higher free energy than the former one. Therefore the
solution with bridge is thermodynamically stable whereas the solution
without bridge is metastable. For 
the solution without a bridge the distortion of the liquidlike layer
around one sphere due to the presence of the other sphere is not visible. Finally,
Fig.~\ref{f:example1}(c) displays the wetting film morphology for the
stable state with bridge at the temperature $T^*=1.2$, i.e., below the
thin-thick transition temperature $T_{tt}^*$. (We note that the
thin-thick transition temperature $T_{tt}$ for each sphere is slightly shifted
by the presence of the second sphere. However, as already pointed out
in Ref.~\cite{dobbsyeomans}, this effect is negligibly small.) In
any case, the difference between the nonlocal and the
local theory is very small. This latter result is in accordance
with the findings for the comparison between the nonlocal and the local
description of the three-phase
contact line on a homogeneous substrate and of the wetting layer
morphology on a chemically structured substrate (compare
Ref.~\cite{bauer1}). For this reason, henceforth we only consider
the local theory.

\begin{figure} 
\begin{center}
\epsfig{file=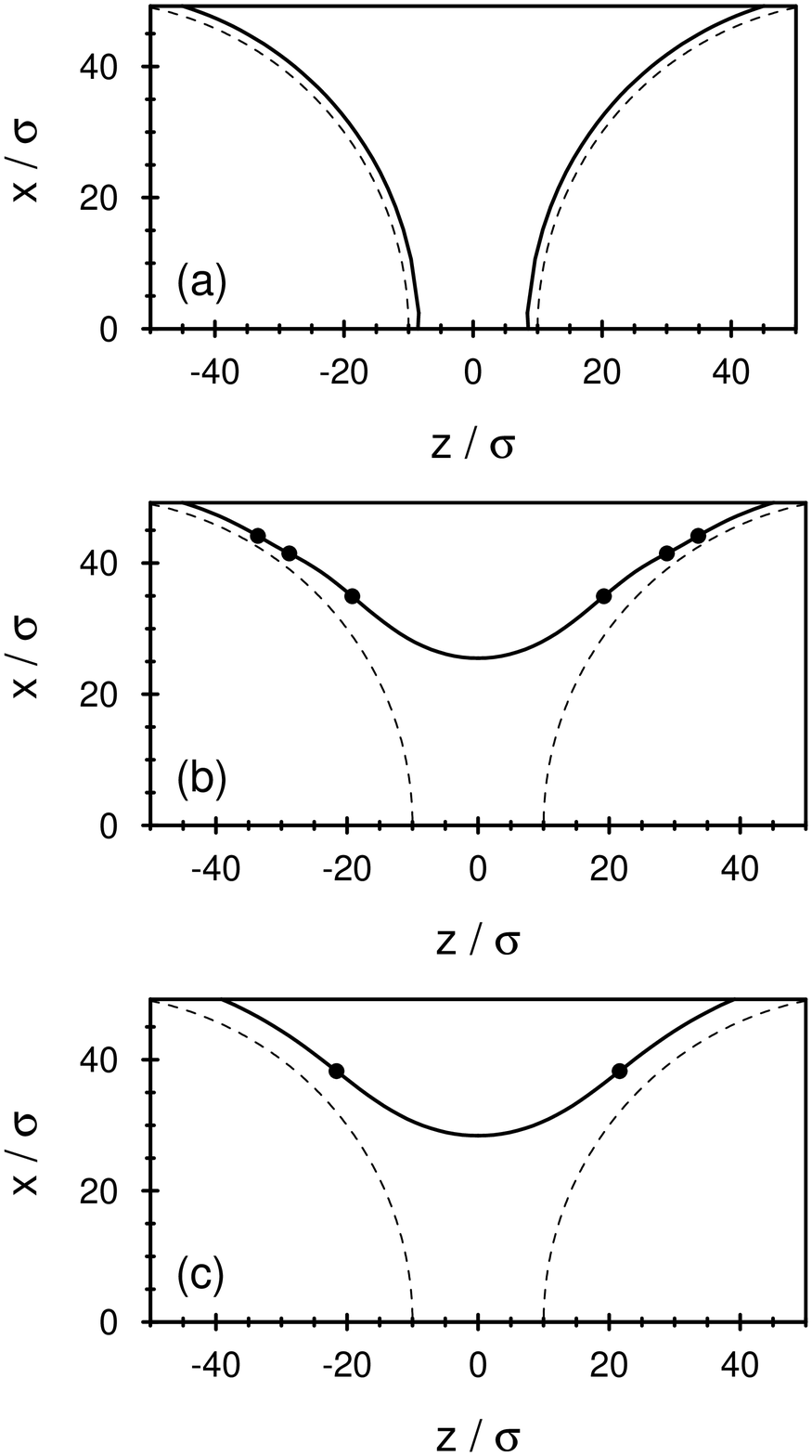, width=7cm, bbllx = 15, bblly = 15,
  bburx = 495, bbury = 840}
\end{center}
\caption{\label{f:example2}
Morphologies of liquidlike wetting layers on two adjacent, identical
spheres with radius $R=50\sigma$, $D=120\sigma$, and for the same
choice of interaction parameters as in Fig.~\ref{f:example1}. The
thick full lines denote the liquid-vapor interface, the thin dashed lines the
surfaces of the spheres. These pictures magnify the
region between the spheres. The temperature is $T^*=1.2$, which 
is above the thin-thick transition temperature $T_{tt}^*\approx1.191$ at
coexistence for a single sphere,
and the pictures differ with respect to the undersaturation:
$\Delta\mu^*=\Delta\mu/\epsilon = 0.05$ in (a), $0.015$ in (b), and
$0.01$ in (c). Between (a) and (b), at $\Delta\mu_{bt}^*\approx
0.0235$ the system undergoes the first-order transition 
from the state without bridge to the state with bridge, and at
$\Delta\mu_{tt}^*\approx 0.0103$ between (b) and
(c) there is a thin-thick transition of the wetting layer around the
single spheres which is the remnant of the prewetting transition on
the corresponding flat
substrate (see, c.f., Fig.~\ref{f:btandttt}). Note that in (b) there
are six turning points ($\bullet$) of the 
profile $h(z)$ whereas in (c) there are only two.}
\end{figure}

Figure~\ref{f:example2} shows another pertinent example. Here we study
the wetting layer morphology for two larger spheres of radius $R=50\sigma$ as
a function of the undersaturation $\Delta\mu$ along the isotherm
$T^*=1.2$. The interaction potential parameters are the same as for
the previous first example and the separation of the surfaces $a$ is
$20\sigma$ ($D=120\sigma$). At coexistence each single
sphere exhibits a first-order thin-thick transition 
at $T_{tt}^* \approx 1.191$ (i.e., $T_{tt}/T_c \approx 0.84$ and
$T_{tt}/T_w \approx 1.13$). In analogy to the prewetting line on a
homogeneous substrate there is a line of thin-thick transitions
$(T,\Delta\mu_{tt}(T))$ which intersects the liquid-vapor coexistence
line at $(T=T_{tt},\Delta\mu=0)$ (compare with, c.f.,
Fig.~\ref{f:btandttt} and Fig.~8(a) in 
Ref.~\cite{bieker}). At the temperature $T^*=1.2>T_{tt}^*$ considered
here the thin-thick transition occurs at
$\Delta\mu_{tt}^* = \Delta\mu_{tt}/\epsilon \approx 0.0103$. Upon reducing the
undersaturation along the isotherm, starting at, e.g., $\Delta\mu^*
= 0.05$, first the 
configuration with thin films and without bridge is stable
(Fig.~\ref{f:example2}(a)). For 
$\Delta\mu\leq\Delta\mu_{bt}$ (\emph{b}ridging \emph{t}ransition) with
$\Delta\mu_{bt}^*\approx 0.0235 > \Delta\mu_{tt}^*(T)$ the solution
with bridge becomes 
stable, but the layers enclosing the spheres still remain thin
(Fig.~\ref{f:example2}(b)). Upon
further reduction of $\Delta\mu$, at $\Delta\mu_{tt}(T)$ 
the second transition from a solution with bridge and thin films to
a solution with bridge and thick films (Fig.~\ref{f:example2}(c))
takes place. (As before, concerning the value of $T_{tt}^*$ at
coexistence, also the value $\Delta\mu_{tt}^*(T)$ is practically
unchanged by the presence of the second sphere -- even for the bridge
configuration.) We note that for this 
choice of parameters and in the case of a solution with bridge and \emph{thin}
films (Fig.~\ref{f:example2}(b)) the profile $h(z)$ exhibits
\emph{six} turning points instead of only two as for the case of a solution with
bridge and \emph{thick} films (Fig.~\ref{f:example2}(c)). This rich curvature
behavior is caused by the details of the 
effective interaction potential between the spherical substrate surfaces and the
liquid-vapor interface (see Sec.~2.3 in
Ref.~\cite{bieker}), similar to the curvature behavior of the 
liquid-vapor interface when it meets a homogeneous, planar substrate
forming a three-phase contact line (compare Ref.~\cite{bauer1}). These features
may also occur for a bridge configuration with thin films at
coexistence and $T<T_{tt}$. 

\subsection{Phase diagram}

\begin{figure} 
\begin{center}
\rotatebox{-90}{\epsfig{file=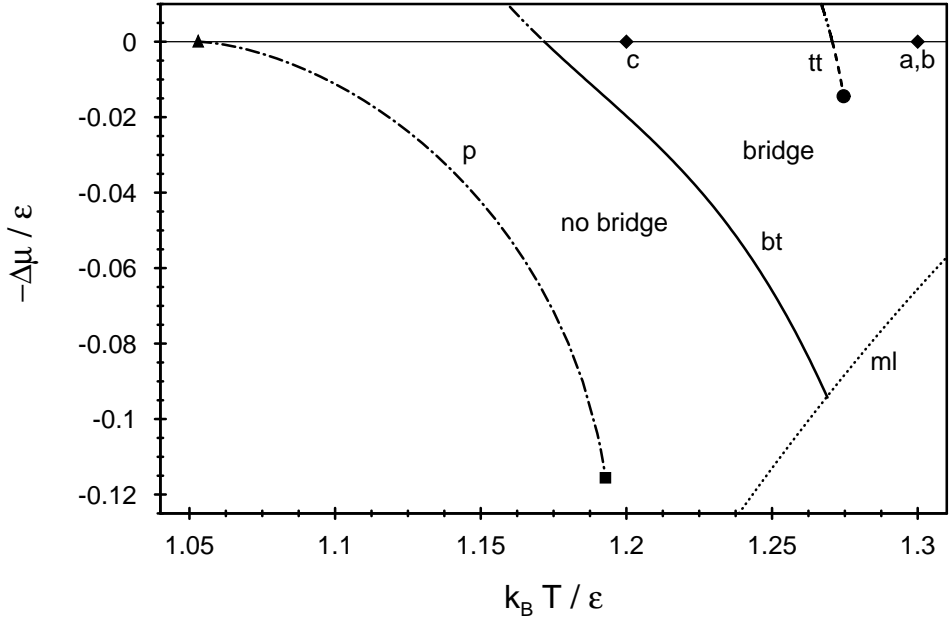, height=13cm, bbllx =
  35, bblly = 40, bburx = 540, bbury = 800}}
\end{center}
\caption{\label{f:btandttt}
Temperature-undersaturation phase diagram of wetting layer
configurations for two spheres with $R=20\sigma$ at a \emph{fixed} distance
$D=50\sigma$ ($a=10\sigma$). The interaction potential parameters are
the same as in Fig.~\ref{f:example1}. The three configurations shown
in Fig.~\ref{f:example1}(a)-(c) are located at the respective
thermodynamic states
``a'' to ``c'' ($\blacklozenge$). The line of liquid-vapor coexistence
$\Delta\mu=0$ separates the region where in the bulk the vapor phase
is stable and the liquid phase is metastable ($-\Delta\mu<0$) from the
region where the liquid phase is stable and the vapor phase is
metastable ($-\Delta\mu>0$). The dotted ``metastability line''
(``ml'') separates the region where the liquid phase in the bulk is
still metastable $(-\Delta\mu>-\Delta\mu_{ml}(T))$ from the region where
only the vapor phase is stable in the bulk $(-\Delta\mu<-\Delta\mu_{ml}(T))$.
The liquidlike layer on each individual sphere exhibits a first-order
thin-thick transition at $-\Delta\mu=-\Delta\mu_{tt}(T)$ (dashed line
``tt''). This line intersects the liquid-vapor coexistence line at
$T_{tt}^*\approx1.271$ and ends at a critical point
($\bullet$) in the vapor phase region; $T_{tt,c}^*\approx1.275$ and
$-\Delta\mu_{tt,c}^*\approx-0.0144$. For the present choice of
interaction potential parameters, at lower temperatures and larger
undersaturations $-\Delta\mu = 
-\Delta\mu_{bt}(T)$ (full line ``bt'') the first-order bridging
transitions between the 
configurations with bridge $(-\Delta\mu>-\Delta\mu_{bt}(T))$ and
without bridge $(-\Delta\mu<-\Delta\mu_{bt}(T))$ occur. This line
intersects the coexistence line linearly. Within the sharp-kink
approximation the line of bridging transitions happens to be cut off
by the ``metastability line''; within a more sophisticated approach
the line ``bt'' is expected to end at a critical point, too. The
locations of the thin-thick transitions in the 
phase diagram are practically not affected by the presence of the
bridge. The dashed-double-dotted lines ($-\cdot\cdot-$) are metastable
extensions of the thin-thick and bridging transition lines, respectively. The
dashed-dotted line ``p'' ($-\cdot-$) is the prewetting line for the
corresponding \emph{planar} substrate. It joins the liquid-vapor coexistence
line $\Delta\mu=0$ tangentially at the first-order wetting transition
temperature $T_w^*\approx 1.053$ ($\blacktriangle$) and ends at a
critical point ($\blacksquare$) in the vapor phase region. For a
discussion of the effects of fluctuations on this mean-field phase
diagram see the main text.}
\end{figure}

The example presented in the previous paragraph shows that besides
the gas-liquid coexistence curve $\Delta\mu=0$ the
$T$-$\Delta\mu$ phase diagram of the system contains two distinct
lines of first-order phase 
transitions: a line of thin-thick transitions $(T,\Delta\mu_{tt}(T))$
on the single spheres (which is the remnant of the line of prewetting
transitions on the corresponding flat substrate and which is, as
stated above, practically unshifted by the presence of the second
sphere) and a second, \emph{independent} line of bridging transitions
$(T,\Delta\mu_{bt}(T))$. If one crosses the latter along an isotherm $T=T_0$
approaching coexistence $(T_0,\Delta\mu\to0)$, at $\Delta\mu=\Delta\mu_{bt}(T_0)$
a transition from the configuration without bridge
$(\Delta\mu>\Delta\mu_{bt}(T_0))$ to a configuration with bridge
$(\Delta\mu<\Delta\mu_{bt}(T_0))$ occurs. The derivative
$\partial\Omega_S/\partial\Delta\mu$ is discontinuous at
$\Delta\mu_{bt}$, indicating that the bridging transition is first
order. Figure~\ref{f:btandttt} shows the 
$T$-$\Delta\mu$ phase diagram for the two spheres with $R=20\sigma$
for $D=50\sigma$ ($a=10\sigma$). The line of thin-thick transitions intersects the
liquid-vapor coexistence line at $T_{tt}^*\approx 1.271$ with a
finite, negative slope (compare Fig.~8(a) in
Ref.~\cite{bieker}). It extends into the vapor phase region
($\Delta\mu>0$) of the phase diagram and ends at a critical point. The
line of bridging transitions intersects the coexistence line also with
a finite, negative slope. On the other end, within our sharp-kink
interface model, it happens to be cut off at that metastability line
in the phase diagram at which the second minimum of 
the bulk free energy at high fluid density (Eq.~(\ref{e:bfedensity})) ceases to
exist so that for larger undersaturations the liquid phase is not even
metastable. Within a more sophisticated approach, e.g., by seeking the
full minimal density distributions of Eq.~(\ref{e:ldafunc}), the line
of bridging transitions is expected to end in a critical point,
too. (Concerning the effect of fluctuations on these mean field
predictions see the following paragraph.) The line of bridging
transitions is entirely located in the region 
where the liquidlike films on the spheres are thin. Moreover, the
effect of the presence of the liquid bridge on the line of thin-thick
transitions is negligibly small. In Fig.~\ref{f:btandttt} the relative
location of the bridging transition line and the thin-thick transition
line corresponds to our specific choice of the interaction potential
parameters as well as the chosen size of and distance between the
spheres. Changing these parameters will lead to shifts of these lines
and, possibly, to different topologies of the phase diagram. Here
we refrain from exhaustingly presenting all possibilities which can
occur according to Refs.~\cite{dobbsetal} and \cite{dobbsyeomans}.

Since the liquid volume enclosed by the interface $h(z)$ is
quasi-zerodimensional, fluctuation effects destroy the sharp
first-order phase transition (see
Refs.~\cite{privmanfisher} and \cite{gelfandlipowsky}). In Sec.~4 of
Ref.~\cite{bieker} it has been extensively discussed how finite size
effects smear out the thin-thick transition such that the thickness
increases sharply but continuously within a range $\delta\mu$ around
$\Delta\mu_{tt}(T)$; these results apply analogously to the present case.
Using similar approximations we obtain a range $\delta\mu$
between $\delta\mu^*\approx0.004$ for $T^*=1.16$ and
$\delta\mu^*\approx0.02$ for $T^*=1.26$
over which the \emph{bridging} transitions shown in
Fig.~\ref{f:btandttt} are smeared out around $\Delta\mu_{bt}(T)$. Thus
close to $\Delta\mu=0$ the quasi-first-order thin-thick transitions
are clearly visible. However, for larger values of $\Delta\mu$ they
become progressively smeared out such that their critical points
predicted by mean field theory are erased by fluctuations.

\section{Effective film-induced interaction potential}
\label{s:eip}

\subsection{Shape of the effective potential, metastability, and
  asymptotic behavior}

In the following we change our point of view: we vary the
distance $D$ between the centers of mass of the spheres instead of
the thermodynamic parameters $T$ and $\Delta\mu$.
Figure~\ref{f:eipforexample1} shows the grand canonical potential
$\Omega_S$ corresponding to the wetting layer morphologies for the
case $R=20\sigma$ and $T^*=1.2$ (Fig.~\ref{f:example1}(c)) as a
function of the separation $a = D-2R$ for several values of
$\Delta\mu$. $\Omega_S$ is the 
minimum of $\Omega_S[h(z)]$ (Eq.~(\ref{e:decomp})) for the given set
of parameters $T$, $\Delta\mu$, and $D = 2R+a$. For each value of
$\Delta\mu$ there are two branches of the free energy, one
corresponding to the solution without bridge, which for the case
$R=20\sigma$ considered here exists only for
$a\gtrsim 0.15R$, and the other corresponding to the solution with 
bridge which exists up to $a\approx 0.65R$ and $a\approx
0.6R$ for $\Delta\mu^*=0$ and $\Delta\mu^*=0.01$, respectively.
At a certain value $D=D_{bt}$ or, equivalently, $a=a_{bt}$, which are
functions of $\Delta\mu$, a 
first-order phase transition occurs with discontinuous derivative
$\partial\Omega_S/\partial D$ between the solutions with and without
bridge. The main effect of increasing the undersaturation
$\Delta\mu$ is that the free-energy curves are rigidly shifted upwards.
This shift is approximately proportional to $\Delta\mu$ and larger in the
case of the solution with bridge, resulting in the dependence of $D_{bt}$ on
$\Delta\mu$. The values of $\Omega_S$ shown in
Fig.~\ref{f:eipforexample1} are obtained within the local theory. The
nonlocal theory yields the same functional dependence
$\Omega_S(D)$ but with a slight and rigid shift of the free-energy
curves, relative to the results of the local theory,
of the order of $0.1\%$ and of the same sign and size for both
the solutions with and without bridge. Finite-size effects again
destroy the sharp first-order bridging transition; we obtain a range
$\delta D\sim0.1\sigma$ (corresponging to $\delta D\sim0.005R$) over
which the bridging transitions shown in Fig.~\ref{f:eipforexample1}
are smeared out. 

The thermodynamic states which are located on the metastable branches
of the free energy curves survive during an average lifetime
$\tau\approx\tau_0\exp(\Delta\Omega_S/k_BT)$ where 
$\Delta\Omega_S$ is the height of the energy barrier that separates
the metastable from the stable branch and $\tau_0$ is a characteristic
microscopic time scale for the dynamics associated with the transition
from a metastable to a stable wetting layer configuration. The energy
barrier is highest 
in the vicinity of the bridging transition and vanishes near the ends
of the metastable branches. An estimation of the energy barrier
height yields, e.g., $\Delta\Omega_S\approx75\epsilon$ for
$\Delta\mu=0$ and $D=50\sigma$ ($a=0.5R$), and with
$k_BT\sim\epsilon$ it follows that $\exp(\Delta\Omega_S/k_BT)\sim
10^{32}$, i.e., the metastable unbridged state for $a=0.5R$ near the
bridging transition remains stable practically forever. However, at, e.g.,
$a=0.2R$ one has $\exp(\Delta\Omega_S/k_BT)\sim10^{11}$ so that
with $\tau_0\sim 1$ps$\dots 1$ns one may observe a decay of the
metastable states near the ends of the metastable branches within
seconds or minutes. Thus the change of the morphology of the wetting
films is expected to exhibit pronounced hysteresis effects as function
of $D$.

Obviously, in the limit of large separation $D\to\infty$ (in which only the
configuration without a bridge is stable) the grand canonical
potential $\Omega_S(D)$ approaches the 
limiting value $2\Omega_S^{(1)}$ corresponding to the free energy of
two individual spheres, each surrounded by a wetting layer.
It is convenient to separate this constant contribution
$2\Omega_S^{(1)}$ from the grand canonical potential $\Omega_S$ of the
system and thus to define an \emph{e}xcess free energy
$\Omega_E(D) = \Omega_S(D)-2\Omega_S^{(1)}$ which contains all
contributions from the wetting-layer induced interaction between the two spheres.
In the limit $D\to\infty$, i.e., in the absence of a liquid bridge,
this excess free energy $\Omega_E(D)$ decays as
$D^{-6}$ (see Eq.~(\ref{e:omegasl_asymp}) and, c.f.,
Sec.~\ref{s:discussion}). We note that for the example 
shown in Fig.~\ref{f:eipforexample1} the coefficient of this leading
order is \emph{positive}, i.e., the effective potential in the
absence of a liquid bridge is \emph{repulsive}. This is owed to the
choice $T<T_{tt}$ for this example: the spheres disfavor the
adsorption of thick liquid 
films and the presence of the second sphere with its surrounding liquidlike
layer leads to an additional cost in free energy which diminishes for
increasing $D$. For the choice $T>T_{tt}$, i.e., if the spheres favor
the adsorption of liquid (e.g., for $T^*=1.3$ as in
Figs.~\ref{f:example1}(a) and (b)) the coefficient of $D^{-6}$ is
\emph{negative} and the effective interaction is \emph{attractive}.
However, in the presence of a liquid bridge, i.e., for sufficiently
small values of $D$, the effective potential
shows the same qualitative behavior as in Fig.~\ref{f:eipforexample1} also for the
case of thick wetting layers ($T^*=1.3>T_{tt}^*$) as well as for the larger
spheres ($R=50\sigma$) with thin or thick films.

\subsection{Effective interaction potential for large spheres}
\label{ss:eipforlargespheres}

\begin{figure} 
\begin{center}
\epsfig{file=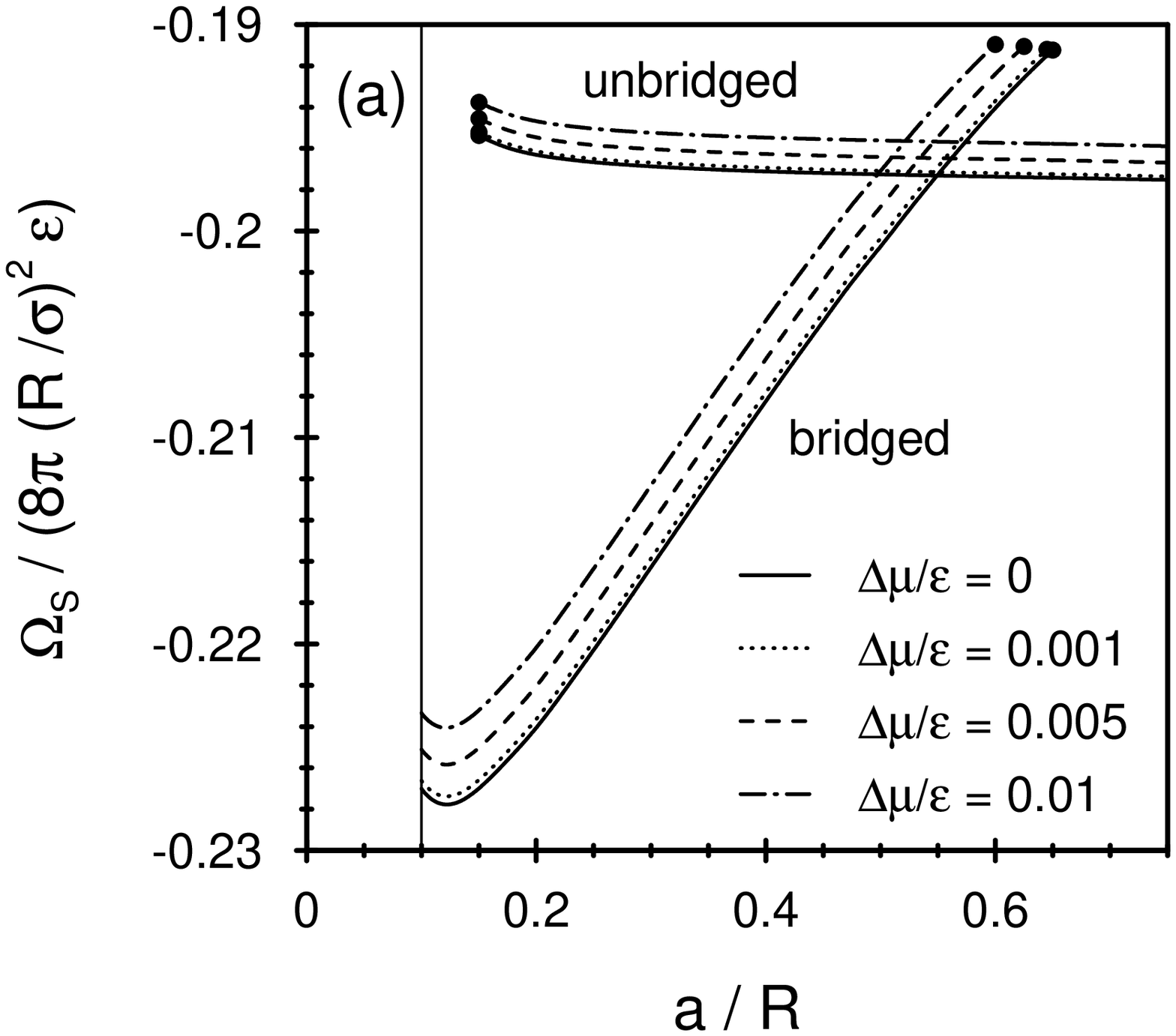, width=7cm, bbllx = 20, bblly = 335,
  bburx = 520, bbury = 775}
\epsfig{file=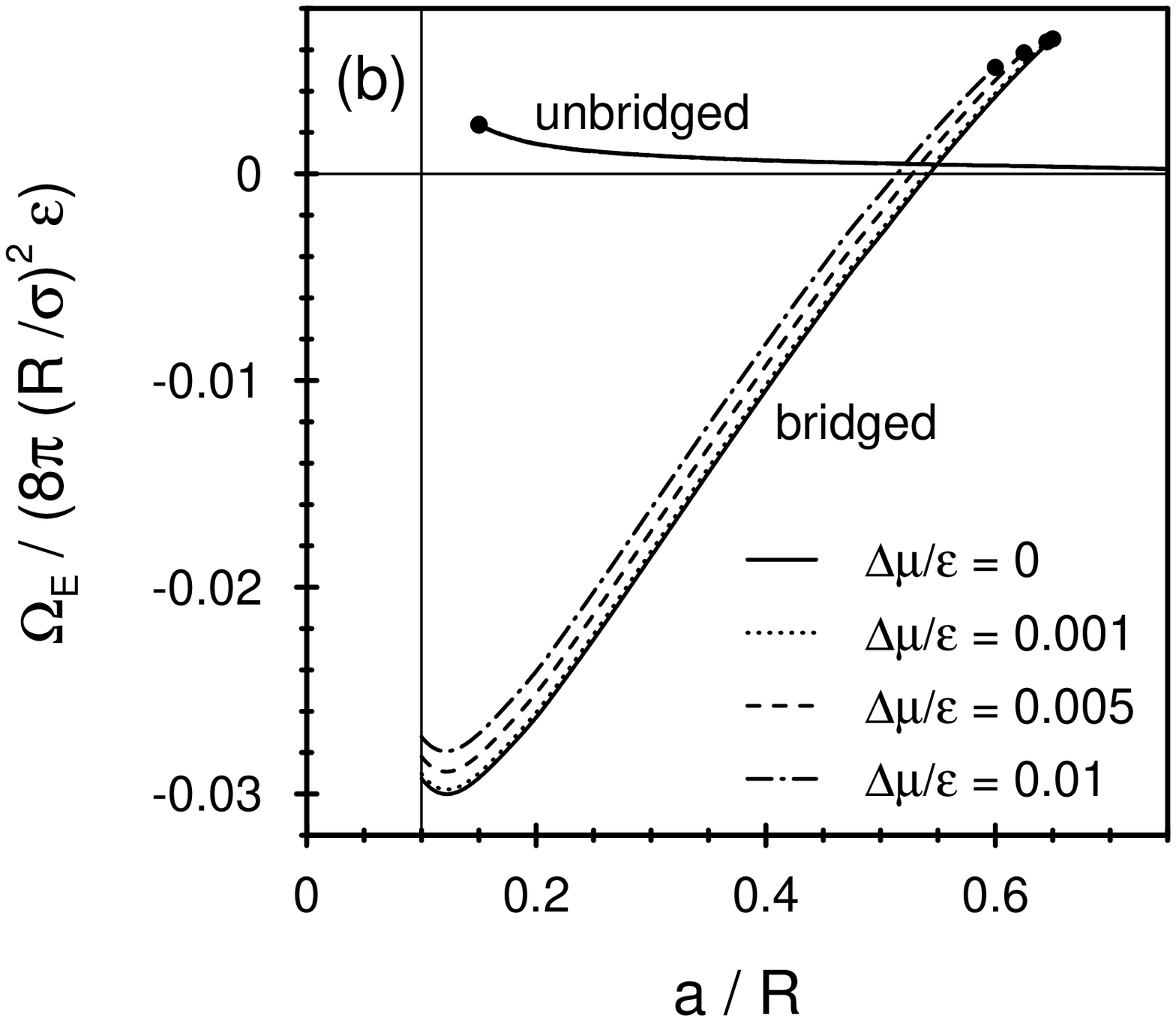, width=7cm, bbllx = 20, bblly = 335,
  bburx = 520, bbury = 775}
\end{center}
\caption{\label{f:eipforexample1}
(a) Dependence of the grand canonical potential $\Omega_S$ on the
separation $a=D-2R$ and the undersaturation $\Delta\mu$ for
the same system as in Figs.~\ref{f:example1}(c) and \ref{f:btandttt},
i.e., for $R=20\sigma$ and 
$T^*=1.2<T_{tt}^*$. The dots indicate the end points of metastable 
branches. For $a<2d_s$ (with $d_s=\sigma=0.05R$ here) the excluded volumes
around the spheres overlap. In the limit $D\to\infty$ the
stable solution is the one without a liquid bridge; in this limit
$\Omega_S(D\to\infty)=2\Omega_S^{(1)}$ is twice the free energy of a single sphere
surrounded by a wetting layer. At the separation $D_{bt}$ or $a_{bt}$, where the two
free energy branches intersect for a given $\Delta\mu$, a first-order
morphological phase transition between a configuration with a liquid bridge
and a state without bridge takes place. The equilibrium thickness
of the homogeneous wetting layer around a single sphere is $l_0
\approx1.3\sigma$, so that $D_{bt}/(R+l_0)\approx2.39$; the slight deviation
from the prediction of Eqs.~(\ref{e:Dbt}) and (\ref{e:lambda}) is due
to the still rather small size of the spheres. We note that, in contrast to
the case shown here, for $T>T_{tt}$
the free energy curve corresponding to the solutions without bridge
approaches its asymptote from \emph{below}. (b) Same as in (a), showing
the excess free energy $\Omega_E = \Omega_S - 2\Omega_S^{(1)}$.
In this presentation the results for the solutions without bridge and
for different undersaturations $\Delta\mu$ collapse
onto a single line. $\Omega_E(D\to\infty)$ decays as $D^{-6}$.} 
\end{figure}

In this subsection we consider the limiting case that the sphere
radius $R$ is much larger than the diameter $\sigma$ of the solvent
particles and that the separations $a$ between the surfaces of the
spheres are proportional to $R$: $R\gg\sigma$, $\sigma\ll a \approx
R$. For such large separations as compared to $\sigma$ the
contributions $\Omega_{sl}$ (Eq.~(\ref{e:omegasl})), $\Omega_{ei}$
(Eq.~(\ref{e:omegaei})), and $\Phi$ (c.f., Eq.~(\ref{e:vdwbetwsph}))
become vanishingly small relative to the contributions $\Omega_{lg}$
(Eq.~(\ref{e:omegalg})) and $\Omega_{ex}$ (Eq.~(\ref{e:excess})). For
the case described above $\Omega_{lg}$ and $\Omega_{ex}$ scale
proportional to the surface area of the spheres, i.e., $\sim R^2$,
whereas for $a/\sigma\to\infty$, $R/\sigma\to\infty$, $a/R$ finite,
$\Phi(D;R)$ remains finite $\sim\epsilon_{ss}\sigma_{ss}^6\rho_s^2$
with a proportionality constant of the order $1$. Analogously, in the
same limit $\Omega_{ei}-2\Omega_{ei}^{(1)}$ (Eq.~(\ref{e:omegaei}))
and $\Omega_{sl}-2\Omega_{sl}^{(1)}$ (Eq.~(\ref{e:omegasl})) are
determined by finite terms \mbox{$\sim\Delta\rho\rho_l\epsilon\sigma^6$} and
\mbox{$\sim\Delta\rho\rho_s\epsilon_{sf}\sigma_{sf}^6$} and of terms
\mbox{$\sim\rho_l^2\epsilon\sigma^6$} and
\mbox{$\sim\rho_l\rho_s\epsilon_{sf}\sigma_{sf}^6$}, respectively, each with
a proportionality constant of the order $1$. Therefore measured in
units of $8\pi R^2$ the unbridged branch of $\Omega_E$ in
Fig.~\ref{f:eipforexample1}(b) vanishes in the limit
$R\to\infty$. Moreover, on this scale the excluded  
volume at small $a$ disappears from the figure, too, because $d_s/R\to0$. 

\begin{figure} 
\begin{center}
\epsfig{file=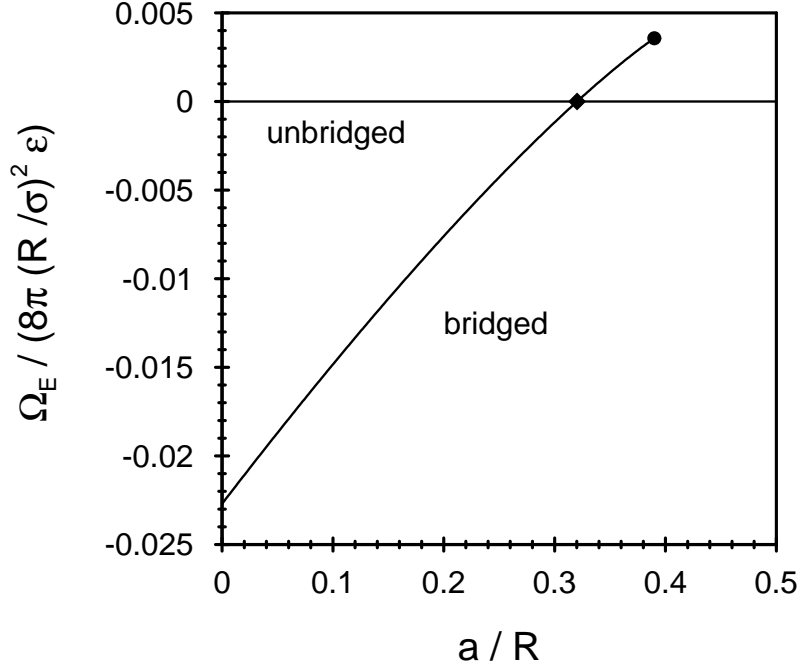, width=10cm, bbllx = 40, bblly = 335,
  bburx = 520, bbury = 775}
\end{center}
\caption{\label{f:eip_largespheres}
Excess free energy $\Omega_E=\Omega_S-2\Omega_S^{(1)}$ for
$\Delta\mu=0$ in the limit of large 
spheres, i.e., $R\gg\sigma$, $\sigma\ll a\approx R$. In this limit the
excess free energy branch for the unbridged solution vanishes if
it is measured in units of $8\pi R^2$. Off two-phase coexistence, i.e.,
for $\Delta\mu\neq0$ the branch for the
bridged solution is determined only by the contributions $\Omega_{lg}$
(Eq.~(\ref{e:omegalg})) and $\Omega_{ex}$ (Eq.~(\ref{e:excess})) to
the free energy. At two-phase coexistence $\Delta\mu$ and
$\Omega_{ex}$ vanish so that $\Omega_E$ is solely determined by
$\Omega_{lg}$. Therefore within the local theory with
$\Omega_{lg}^{(loc)}$ (Eq.~(\ref{e:local})) the bridged solution is a
minimal area surface, i.e., its mean curvature is zero. Since $a\gg
d_s\approx\sigma$ the excluded 
volume at small $a$ disappears from the figure. Therefore, compared
with the full curve in Fig.~\ref{f:eipforexample1}(b) the potential
curve here is effectively shifted to smaller values of $a$. Moreover,
the actual minimum of the effective interaction potential at small
$a\gtrsim\sigma$ (compare
Fig.~\ref{f:eipforexample1}), which is due to the influence of the
contributions $\Omega_{ei}$ and $\Omega_{sl}$, is not visible on this
scale either. The critical separation
for the bridging transition ($\blacklozenge$) is given by $a_{bt}/R\approx
0.32$ (Eqs.~(\ref{e:Dbt}) and (\ref{e:lambda})). If the thermodynamic
state of the system is driven into the off-coexistence region
$\Delta\mu>0$ the whole excess free energy branch for the bridged
solution is shifted upwards (compare Fig.~\ref{f:eipforexample1}). For any
finite value of $\Delta\mu$, in the limit $R\to\infty$ there is no
bridging transition anymore (see the main text).}
\end{figure}

Fig.~\ref{f:eip_largespheres} shows the excess effective interaction
potential $\Omega_E$ in the limit of large spheres for the case
$\Delta\mu=0$, i.e., at two-phase coexistence in the solvent. In this
limit and for $\Delta\mu=0$, $\Omega_{lg}$ is the only 
relevant contribution to $\Omega_S$ because
$\Omega_{ex}(\Delta\mu=0)=0$. Accordingly, in this case the bridging
transition is determined by the equality of the surface areas of the
liquid-vapor interfaces for the unbridged and bridged
configuration. From this condition and from dimensional analysis it
follows that for large spheres $D_{bt}(\Delta\mu=0)$ is determined by
the equation
\begin{equation}\label{e:surfaceequality}
8\pi(R+l_0)^2\sigma_{lg}^{(p)} = 8\pi(R+l_0)^2\sigma_{lg}^{(p)}
f\left(\frac{D_{bt}}{R+l_0}\right) 
\end{equation}
where $f$ is, for
dimensional reasons, a universal function of $D/(R+l_0)$
alone which describes the surface area of the bridged
configuration; $l_0$ is the equilibrium wetting layer thickness on a
single sphere. Since the line of bridging transitions lies below the
line of thin-thick transitions, $l_0$ remains microscopicly small at
the bridging transition (Fig.~\ref{f:btandttt}). Therefore one has 
\begin{equation}\label{e:Dbt}
D_{bt}(\Delta\mu=0) = \lambda(R+l_0)
\end{equation}
with a universal number 
\begin{equation}\label{e:lambda}
\lambda\approx2.32
\end{equation}
determined by $f(\lambda)=1$ (compare
Fig.~\ref{f:eip_largespheres}). If one applies this reasoning to 
Fig.~\ref{f:eipforexample1} one finds $\lambda\approx2.39$. Therefore
even for $R=20\sigma$ this macroscopic approximation leads to a
surprisingly small error of only $3\%$ for
$D_{bt}(\Delta\mu=0)$. Accordingly, in Fig.~\ref{f:eipforexample1} the
full curves corresponding to $\Delta\mu=0$ closely resemble the ones
in Fig.~\ref{f:eip_largespheres} describing the case of large
spheres. The only differences appear for small separations $a$ where
the bridged branch linearly extends down to its minimum value
$\Omega_E/(8\pi(R/\sigma)^2) \approx -0.0227\epsilon$ at $a/R=0$
(Fig.~\ref{f:eip_largespheres}). Only in this range of separations the
effect of the contributions $\Omega_{ei}$ and $\Omega_{sl}$ becomes
significant, leading to the deeper minimum visible in
Fig.~\ref{f:eipforexample1}. Thus for $\Delta\mu=0$ and large $R$
the dependence of the effective interaction potential on $R$ for the
bridged configuration is captured by the indicated rescaling of the
axes in Fig.~\ref{f:eipforexample1}(b). However, our numerical
analysis shows that the smallness of the deviations between the
macroscopic description valid for $R\gg\sigma$ and the actual results
for $R=20\sigma$ is somewhat fortuitous. Whereas the dependence of
$D_{bt}(\Delta\mu=0)$ on $R$ is indeed weak, the shape of the
potential (for $\Delta\mu=0$) reduces to that shown in
Fig.~\ref{f:eip_largespheres} only for $R$ larger than several hundred
$\sigma$ and, surprisingly, for $R$ up to $20\dots30\sigma$ with the
deviations being maximal for $R\approx100\sigma$.

Off coexistence $\Delta\Omega_b = \Omega_b(\rho_l)-\Omega_b(\rho_g)
\approx \Delta\mu\Delta\rho$ is positive so that
Eq.~(\ref{e:surfaceequality}) has to be augmented correspondingly:
\begin{equation}
8\pi(R+l_0)^2 +
\frac{8\pi\Delta\Omega_b}{3\sigma_{lg}^{(p)}}\left((R+l_0)^3-R^3\right)
= {\mathcal A}+\frac{\Delta\Omega_b}{\sigma_{lg}^{(p)}}\mbox{Vol}(\mL)
\end{equation}
where ${\mathcal A}$ and $\mbox{Vol}(\mL)$ (Eq.~(\ref{e:volliq})) are
the area of the liquid-vapor interface and the volume of the liquid,
respectively, for the bridged configuration. They are obtained by
inserting into Eqs.~(\ref{e:local}) and (\ref{e:volliq}) that profile
$h(z)$ which solves the differential equation determining the minimum
of $\Omega_{lg}[h]+\Omega_{ex}[h]$ together with the appropriate
boundary conditions. By splitting off a factor $(R+l_0)^2$ from
${\mathcal A}$ and $(R+l_0)^3$ from $\mbox{Vol}(\mL)$ dimensional
analysis shows that up to terms $\sim l_0/R$ the critical distance for
the bridging transition is given by a universal scaling function
$\Lambda$:
\begin{equation}\label{e:Lambda}
D_{bt}(\Delta\mu) =
\Lambda\left(\frac{\Delta\rho\Delta\mu R}{\sigma_{lg}^{(p)}}\right) R
\end{equation}
with $\Lambda(0) = \lambda$. Thus off coexistence the critical
bridging transition depends, apart from an explicit factor $R$, on $R$
and $\Delta\mu$ via the scaling variable
$\Delta\rho\Delta\mu R/\sigma_{lg}^{(p)}$. This property is shared by
the whole bridged branch of the effective interaction potential. Thus
increasing $R$ for fixed undersaturation $\Delta\mu$ has the same
effect as increasing $\Delta\mu$ for fixed $R$. From
Fig.~\ref{f:eipforexample1}(b), in which the unbridged branch will
disappear in the limit $R\gg\sigma$, one infers that the range and the depth 
of $\Omega_E$ decrease for increasing $R$ at fixed undersaturation
$\Delta\mu$. The behavior of $D_{bt}$ and of the bridged branch of the
effective interaction potential off coexistence and for $R\to\infty$
is determined by the behavior of the scaling function $\Lambda(x)$ in the limit
$x\to\infty$. Our numerical data indicate that $\Lambda(x\to\infty)<2$
so that due to the geometric constraint $D\geq2R$ there is no bridging
transition and the bridged branch 
of the effective potential vanishes for any value of $\Delta\mu$ 
in the limit $R\to\infty$. The cost in free energy due to the excess
contribution $\Omega_{ex}$ suppresses the formation of a liquidlike
bridge in the case of macroscopicly large spheres. In turn, this
means that for any finite 
value of $\Delta\mu$ there is a large but finite critical radius $R_c$ for
which the critical separation $a_{bt}$ for the bridging transition
attains the value $a_{bt}=0$, such that for $R>R_c$ there is no
bridging transition. The determination of $R_c$ requires to analyze
the full dependence of 
$\Lambda$ on the scaling variable $x$. This, however, implies such a large
numerical effort that it is beyond the scope of the present paper.

\section{Discussion}
\label{s:discussion}

\subsection{Total interaction potential}

The bare dispersion interaction between the two spheres is not included
in Eq.~(\ref{e:ldafunc}). According to Hamaker~\cite{hamaker} this
contribution is given by
\begin{equation}\label{e:vdwbetwsph}
\Phi(D;R) = -\frac{A_{ss}}{12}\left(\frac{4R^2}{(D-2R)(D+2R)} +
\frac{4R^2}{D^2} + 2\ln\left(\frac{(D-2R)(D+2R)}{D^2}\right)\right)
\end{equation}
as the dispersion interaction between two
identical spheres of radius $R$ at center-of-mass distance $D$. In
the limit $a/R\ll1$, where $a = 
D-2R$ (see Fig.~\ref{f:system}(a)) is the smallest separation between the
surfaces of the spheres, Eq.~(\ref{e:vdwbetwsph}) reduces to
\begin{equation}\label{e:barevdw_smalla}
\Phi(D=2R+a;R\gg a) \approx -\frac{A_{ss}}{12}\,\frac{R}{a}
\end{equation}
which corresponds to the Derjaguin approximation whereas $\Phi(D\gg
R;R)=-16A_{ss}R^6/(9D^6)$. Thus except for the $D$-independent bulk
contribution $\mbox{Vol}(\mV_f)\Omega_b(\rho_g)$
the total grand canonical potential of the system is
\begin{equation}\label{e:omegatot}
\Omega_{tot}(D;R) = \Omega_S(D;R) + \Phi(D;R)
\end{equation}
where $\Omega_S(D;R)$ is given by the minimum value
$\min_{\{h(z)\}}(\Omega_S[h(z)])$ for given $D$ and $R$
(Eqs.~(\ref{e:decomp}) and (\ref{e:gcpismin})); in analogy to
$\Omega_E$ we define the excess total free energy $\Omega_{E,tot} =
\Omega_{tot}-2\Omega_S^{(1)}$. $A_{ss}$ is the Hamaker 
constant appertaining to the bare dispersion interaction between the
particles in the spheres. In the case of pairwise additivity of the molecular  
interactions and in the absence of retardation effects one has $A_{ss} =
4\pi^2\epsilon_{ss}\sigma_{ss}^6\rho_s^2$ if the interaction potential
between two individual molecules in the spheres is given by a
Lennard-Jones potential 
(Eq.~(\ref{e:ljsphere})) with the parameters $\epsilon_{ss}$ and
$\sigma_{ss}$. Typically $A_{ss}$ is of the order of $10^{-19}$J or, equivalently,
$10\dots100\epsilon$. If the vacuum between the spheres is replaced by
a medium of condensed matter the interaction between the
spheres is screened~\cite{israelachvili}. In our present model this medium
is the bulk vapor phase modified by the presence of the liquidlike films
adsorbed on the spheres and the screening effect is
described microscopicly by the functional $\Omega[\rho(\rv)]$.

\begin{figure} 
\begin{center}
\epsfig{file=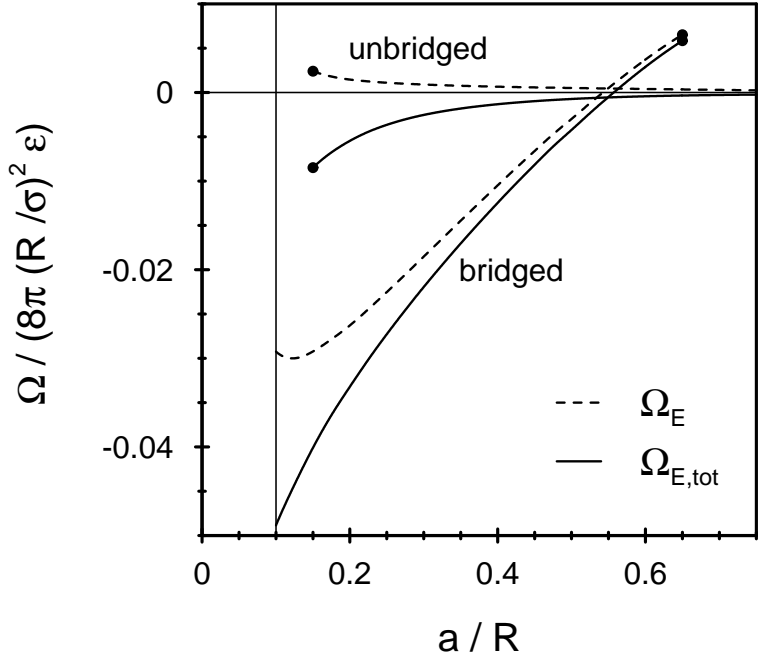, width=10cm, bbllx =
  40, bblly = 335, bburx = 520, bbury = 775}
\end{center}
\caption{\label{f:eipforexample1plusbare}
Excess free energy
$\Omega_E = \Omega_S-2\Omega_S^{(1)}$ (dashed lines) and excess total
free energy $\Omega_{E,tot} =
\Omega_S-2\Omega_S^{(1)}+\Phi$ (full lines)
for $\Delta\mu=0$. Here $T^*=1.2$ and $R=20\sigma$ so that the dashed
lines are identical with 
the full lines in Fig.~\ref{f:eipforexample1}(b). The dots indicate
the end points of metastable branches. The parameters $\epsilon_{ss}$
and $\sigma_{ss}$ of the pair potential between the particles forming
the spheres are chosen such that the condition 
$A_{sf} = \sqrt{A\,A_{ss}}$ for the corresponding Hamaker constants is
satisfied. Although the wetting-layer induced potential for the
solutions without bridge is \emph{repulsive}, the total interaction
potential including the bare dispersion potential is
\emph{attractive}. For small separations $a$ or $D$ the bare dispersion
potential dominates. In the limit $D\to\infty$, i.e., for
the configurations without bridge, $\Omega_E$ and $\Omega_{E,tot}$
decay as $D^{-6}$ as expected for dispersion interactions.}
\end{figure}

In Refs.~\cite{vold} and \cite{vincentetal} this
additional screening effect, due to spherical shells of adsorbed,
homogeneous layers surrounding
spherical particles, on the dispersion interaction between the
latter immersed in another homogeneous medium has been calculated
macroscopicly. Beyond molecular scales these results should closely
correspond to the  
configuration without liquid bridge discussed herein because the
deviation of the spherical shape of one wetting layer due to the
presence of the second sphere is very small. Indeed, the
interaction energy calculated in Refs.~\cite{vold} and 
\cite{vincentetal} is practically the same as the sum of the $D$-dependent
contributions in $\Omega_{ei}$ (Eq.~(\ref{e:omegaei})) and
$\Omega_{sl}$ (Eq.~(\ref{e:omegasl})) for the configurations without
bridge -- for these configurations $\Omega_{lg}$ and $\Omega_{ex}$ do
not contribute to the dependence of $\Omega_S$ on $D$ -- and the direct 
dispersion interaction $\Phi(D;R)$. In Ref.~\cite{vold} the total dispersion 
interaction is shown to be always attractive if the Hamaker constants
$A_{ij}$ corresponding to the interaction between any two media $i$ and
$j$ are chosen such that $A_{ij} = \sqrt{A_{ii}\,A_{jj}}$. Although the
effective interaction induced by the wetting layers shown in
Figs.~\ref{f:eipforexample1}(b) and \ref{f:eipforexample1plusbare} for the
configuration without bridge is repulsive, we note that the sum
$\Omega_{tot}$ of this interaction and of the bare dispersion 
potential $\Phi(D;R)$ is also \emph{attractive} if we choose the Hamaker
constant in Eq.~(\ref{e:vdwbetwsph}) accordingly, i.e., $A_{ss} =
A_{sf}^2/A$ (Fig.~\ref{f:eipforexample1plusbare}). Therefore 
our results are consistent with those obtained in Ref.~\cite{vold}. Since only
effective interactions between finite volumes enter into the total
excess interaction potential $\Omega_{E,tot}$ and these
effective interactions decay as $D^{-6}$ in the limit of large
separations $D$, the same holds for $\Omega_{E,tot}$.

Figure~\ref{f:eipforexample1} shows that as soon as the wetting films
snap to a liquidlike bridge, whether it is stable or metastable, there
is an attractive wetting-layer-induced force
$-\partial\Omega_E/\partial D$ that pulls the 
spheres together. From Fig.~\ref{f:eipforexample1} one can infer that
this attractive force is of the order of 
$40\epsilon/\sigma$ in the range between $a\approx4\sigma$ (i.e.,
$0.2R$ for $R=20\sigma$ discussed in this figure) and $a\approx
10\sigma$ ($0.5R$) where the effective potential varies almost
linearly. At the small separation $a_{min}\approx2.5\sigma$ 
the effective potential $\Omega_E$ induced by
the bridgelike wetting layer is minimal and the wetting-induced force
is zero. Finally, at smaller separations the interaction is repulsive
leading to a stabilization of the spheres at $D=D_{min}=2R+a_{min}$. Within the
range $a\ll R$ the bare, direct
dispersion interaction between the spheres
(Eqs.~(\ref{e:vdwbetwsph}) and (\ref{e:barevdw_smalla})) gives rise to a
force $F_{bare}(a) \approx -A_{ss}R/12a^2$. The estimate
$A_{ss}\approx4\pi^2\epsilon_{sf}^2\sigma_{sf}^{12}\rho_{s}^2/\epsilon\sigma^6 
\sim 400\epsilon$ 
for the case of pairwise additive interactions without retardation
follows from the ansatz $A_{sf} = \sqrt{A\,A_{ss}}$, so that 
the bare dispersion force in our example with $R=20\sigma$ is
$F_{bare}(a)\approx -670\epsilon\sigma/a^2$. Therefore in the range where the
bridge-induced force is almost constant ($4\sigma\lesssim a\lesssim10\sigma$)
the direct, bare dispersion
force decays from approximately $-40\epsilon/\sigma$ (which is of the same order
of magnitude as the bridge-induced force) to approximately $-6\epsilon/\sigma$,
whereas for smaller separations it becomes the dominant force. 

\subsection{Relevance for force microscopy}
\label{ss:forcemicroscopy}

Our model calculations can be tested experimentally by force
microscopy. This can be done by suitably fixing one sphere in the
fluid and by attaching the second one to the tip of a force
microscope. Alternatively, both spheres can be positioned by optical
tweezers and the force law can be inferred by monitoring optically
their dynamics after switching off the tweezers. At separations
between the spheres which are comparable with the diameter $\sigma$ of
the solvent particles the actual effective interaction potential will
exhibit an additional oscillatory contribution
due to packing effects which decays exponentially on the scale
$\sigma$~\cite{kinoshita}. In order to obtain these 
oscillations one would have to resort to density functional theories
which are more sophisticated than the one in
Eq.~(\ref{e:ldafunc}). This, in turn, would make it much more
difficult to obtain the bridgelike configuration, to map out the
complete phase diagram, and to obtain results for large
spheres. According to Subsec.~\ref{ss:eipforlargespheres}, for
$R\gg\sigma$ and at two-phase coexistence $\Delta\mu=0$ the bridging
transition occurs at distances $a$ which are proportional to $R$. In
this case, due to $R\gg\sigma$, the effective interaction potential
will be practically unaffected by this oscillatory contribution for
the vast portion $\sigma\ll a\ll a_{bt}$ of the range of the effective
interaction potential.

\subsection{Relevance for charge stabilized colloidal suspensions}
\label{ss:cscs}

Whereas the kinds of experiments considered in the previous subsection
are focused on two individual
spherical particles, we discussed in the Introduction that the
effective interaction potential enters into the collective behavior of
colloidal suspensions such that the bridging transition may trigger
flocculation. If colloidal suspensions would be governed by dispersion
forces alone, most of them would flocculate even in the absence of the
wetting-induced forces discussed here because the dispersion forces
generate the so-called primary minimum in the effective interaction
potential close to contact. Since this minimum is much deeper than
$k_BT$ the colloidal particles would simply stick together
permanently. This effect, which is undesired for many applications,
can be avoided by endowing the particles with electrical charges which
adds a screened Coulomb repulsion between the charged particles. As
a result, such charge stabilized colloidal suspensions are characterized
by effective interaction potentials in which a substantial energy
barrier separates the aforementioned primary minimum from a second,
much more shallow minimum at larger distances. Since this potential
barrier is typically large compared with $k_BT$ the phase behavior of
the colloidal particles is practically independent of the primary
minimum formed by the dispersion forces and determined by the
shape of the potential \emph{outside} the barrier. As demonstrated by
Figs.~\ref{f:eipforexample1} and \ref{f:eipforexample1plusbare} the
range of the wetting-induced forces is about $0.55R$, in good agreement
with $D_{bt}\approx2.32(R+l_0)$ (see Eqs.~(\ref{e:Dbt}) and (\ref{e:lambda})).
On the other hand the position (and height) of the aforementioned
energy barrier depends sensitively on the size of the total charge on
the spheres, the amount of salt in the solvent, and the dispersion
forces and can be varied over a wide range. With a high salt
concentration the barrier position can be as small as a few tens of
nm. Thus under such circumstances the wetting-induced interaction
potentials would be relevant even for colloidal particles whose radii
are only a few tens of nm.

\subsection{Relevance for stericly stabilized colloidal suspensions}
\label{ss:sscs}

\begin{figure} 
\begin{center}
\epsfig{file=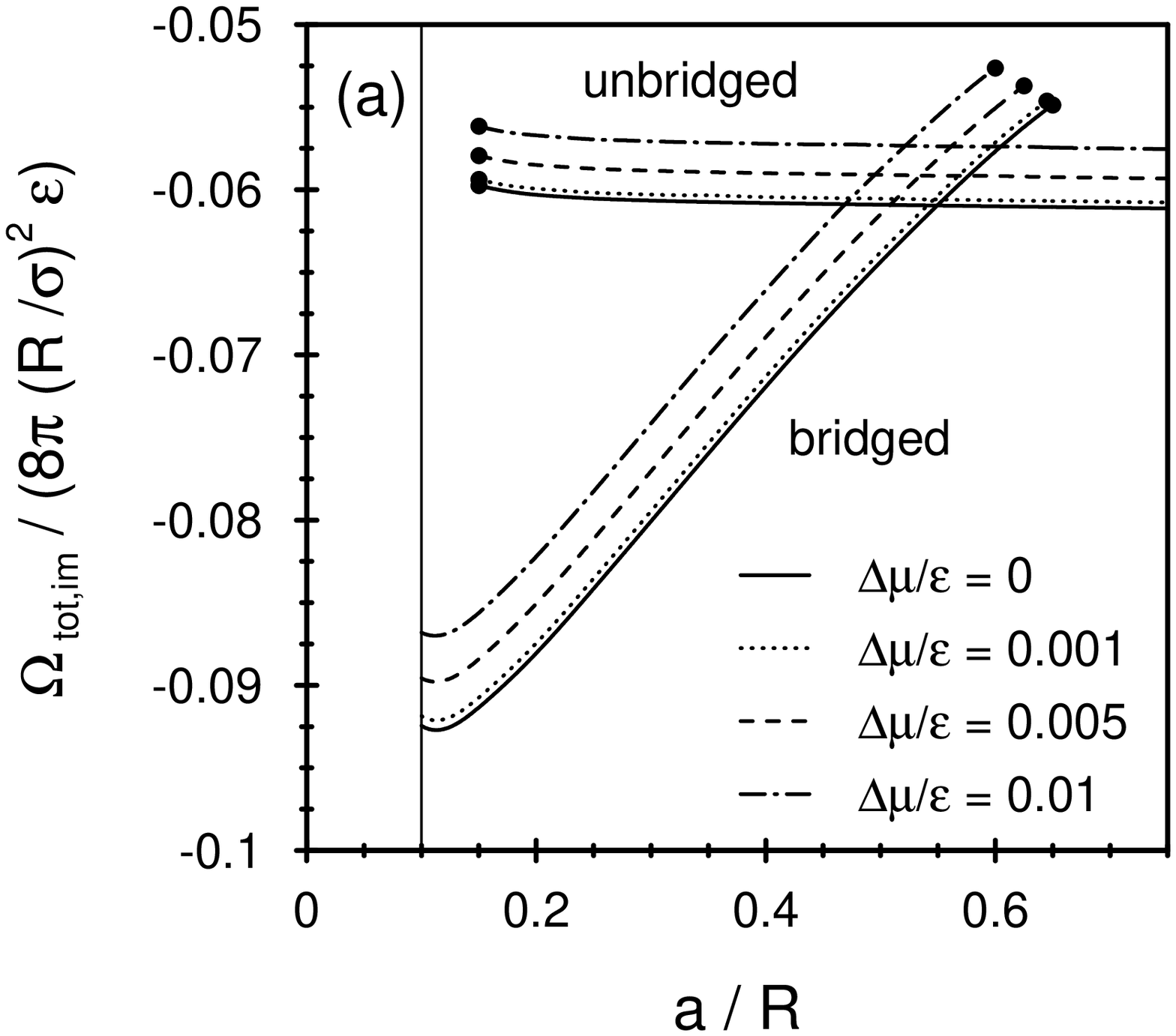, width=7cm, bbllx = 20, bblly = 335,
  bburx = 520, bbury = 775}
\epsfig{file=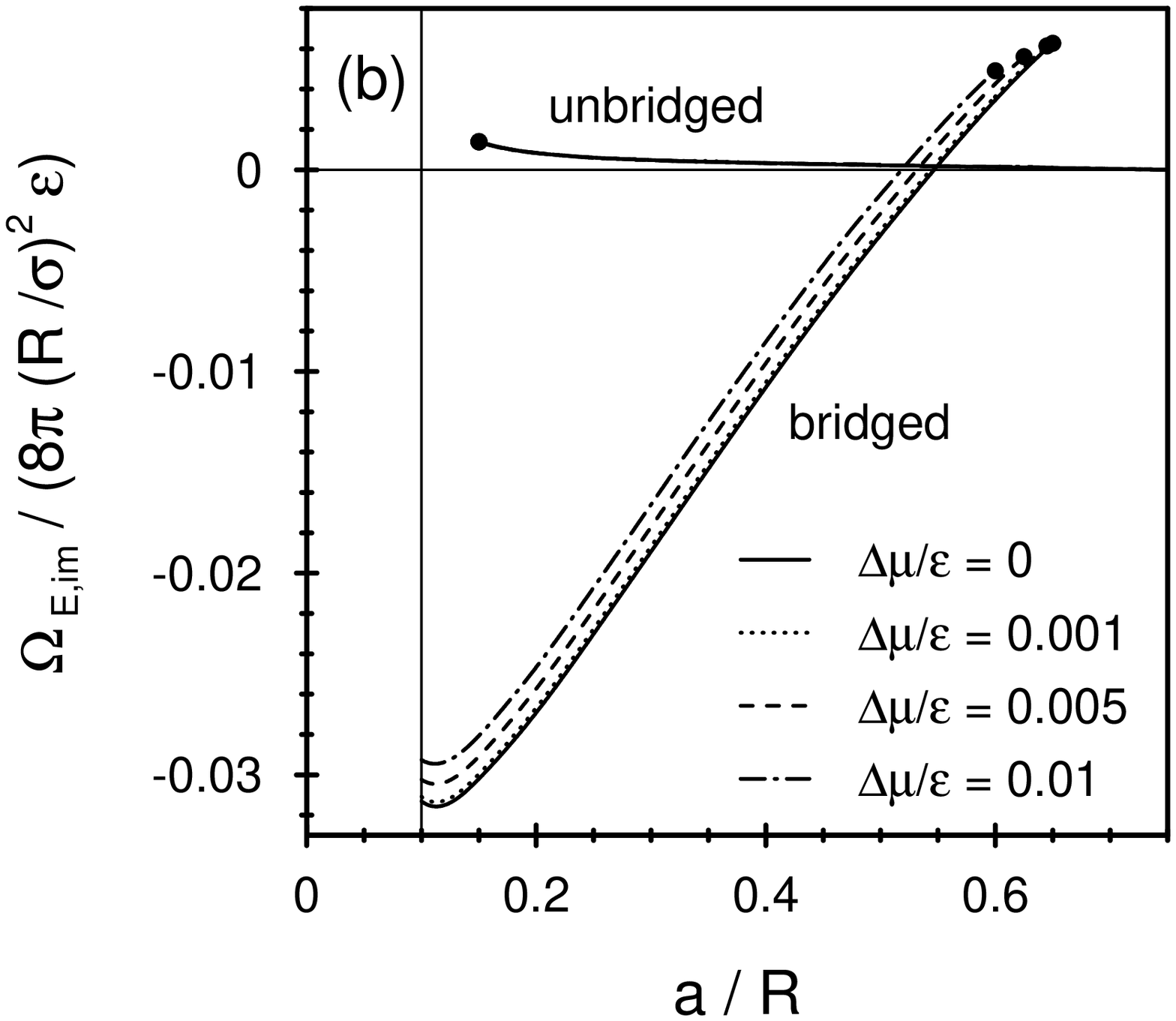, width=7cm, bbllx = 20, bblly = 335,
  bburx = 520, bbury = 775}
\end{center}
\caption{\label{f:indexmatch}
Same as in Fig.~\ref{f:eipforexample1} but with $\Omega_{tot,im} =
\Omega_S-\Omega_{sg}$ (a) and with $\Omega_{E,im} =
\Omega_{tot,im}-2\Omega_{im}^{(1)}$ (b). We choose again $T^*=1.2$, $R=20\sigma$,
and the interaction parameters as in the previous figures. The dots
indicate the end points of metastable branches. The total interaction
potential for index-matched spheres and bulk fluid is again
\emph{repulsive}: since the temperature is below the thin-thick transition
temperature $T_{tt}$ the adjacent spheres dislike the presence of
additional liquid in their vicinity and therefore it is energetically advantageous
to separate them as much as possible. $\Omega_{E,im}$ for the
solutions without bridge is smaller than $\Omega_E$. However, for the
bridged solutions, $\Omega_{E,im}$ and $\Omega_E$ as well as the
corresponding wetting-induced forces are of almost the same size,
respectively.}
\end{figure}

There is another class of colloidal suspensions for which the
wetting-induced forces can be of practical importance. By coating the
colloidal particles with polymers and by matching the refractive
indices of the colloidal particles and the bulk fluid (in our
case study the vapor phase or, more realisticly in the present
context, the A-rich liquid phase of a binary liquid mixture acting as
the solvent) the colloidal particles behave effectively like hard spheres (see,
e.g., Refs.~\cite{puseyvanmegen} and
\cite{poonpuseylekkerkerker}). Through this index matching the sum 
of the bare interaction potential $\Phi(D;R)$ and the effective
interaction potential $\Omega_{sg}$, which would arise if the spheres
were immersed in the homogeneous and unperturbed bulk solvent,
vanishes. Within our model $\Omega_{sg}$ is given by the expression in
Eq.~(\ref{e:omegasl}) with $\rho_l$ replaced by $\rho_g$, which is the
density of the bulk phase. Since the index matching works for the bulk
phase, it does not work for the wetting phase. As a consequence the
wetting-induced forces appear against a background effective potential
of hard spheres. Therefore for this class of colloidal suspensions the
wetting phenomena discussed here are expected to have a pronounced
effect on their phase behavior. Within our model, for
\emph{i}ndex-\emph{m}atched suspensions the \emph{total} effective
interaction potential is given by
\begin{equation}
\Omega_{tot,im}(D;R) = \Omega_{tot}(D;R)-(\Phi(D;R)+\Omega_{sg}(D;R))
= \Omega_S(D;R)-\Omega_{sg}(D;R)
\end{equation}
and in analogy to $\Omega_E$ and $\Omega_{E,tot}$ we define
\begin{equation}
\Omega_{E,im}(D;R) = \Omega_{tot,im}(D;R)-2\Omega_{im}^{(1)}(R)
\end{equation}
with $\Omega_{E,im}(D\to\infty;R) = 0$ for the unbridged
solutions. Figure~\ref{f:indexmatch} displays $\Omega_{tot,im}$ and
$\Omega_{E,im}$ as function of $a=D-2R$ for the same system as in
Figs.~\ref{f:eipforexample1} and \ref{f:eipforexample1plusbare}.
$\Omega_{sg}$ is about $30\%$ smaller than $\Omega_S$ for the unbridged
solution and also approaches its asymptote $2\Omega_{sg}^{(1)}$ from
above. As before (see the discussion of
Fig.~\ref{f:eipforexample1plusbare} above), the resulting total
effective interaction between spheres in an 
index-matched bulk fluid for the state with liquid bridge is still
attractive and of the same order of magnitude as the bare dispersion
interaction between the spheres, i.e., in the absence of the solvent.

\section{Summary}
\label{s:summary}

We have obtained the following main results:
\begin{enumerate}
\item Based on microscopic interaction potentials and within a simple
  version of density functional theory
  (Eqs.~(\ref{e:ldafunc})--(\ref{e:w0})) we have calculated the grand
  canonical potential of a system of two spheres immersed in a bulk
  fluid phase (Fig.~\ref{f:system}). The microscopic interactions are
  chosen such that the spheres
  prefer the adsorption of a second fluid phase which is
  thermodynamically close to the bulk fluid phase. Accordingly, a
  single sphere immersed in the fluid is covered by a homogeneous
  wetting layer of this second phase of thickness $l_0$. These thin wetting
  layers covering the spheres lead to an effective wetting-induced
  interaction potential $\Omega_S(D)$ between the spheres. We have
  systematically determined the dependence of $\Omega_S$ on the
  distance $D$ between the spheres in terms of the morphology
  $h(z)$ of the wetting film enclosing the spheres
  (Eqs.~(\ref{e:decomp})--(\ref{e:omegasl})). We find that the shape
  of the effective interaction potential $\Omega_S(D)$ depends, inter
  alia, on the effective
  interaction of two spheres immersed in the homogeneous \emph{wetting}
  phase (Eq.~(\ref{e:omegasl})). This contribution, which is
  independent of $h(z)$, is not incorporated in previous
  phenomenological models for this system~\cite{dobbsetal,dobbsyeomans}.
\item The equilibrium interfacial profiles of the wetting
  layers are determined numerically by minimizing the free
  energy functional $\Omega_S[h(z)]$ in
  Eqs.~(\ref{e:decomp})--(\ref{e:omegasl}). We have calculated the
  rich structure of these 
  equilibrium profiles (Fig.~\ref{f:example2}) for spheres of radii $R=20\sigma$
  (Fig.~\ref{f:example1}) and $R=50\sigma$ (Fig.~\ref{f:example2})
  where $\sigma$ denotes the diameter of the solvent
  particles. As function of distance $D$, temperature $T$, and
  undersaturation $\Delta\mu$ the system undergoes a first-order
  ``bridging transition'' between the two configurations shown in
  Fig.~\ref{f:system}. For a fixed distance $D$ we have
  mapped out the phase diagram of bridging transitions in the $T$-$\Delta\mu$ plane
  (Fig.~\ref{f:btandttt}). It turns out that the bridging transition
  differs from and to a large extent is independent of the
  thin-thick transition of the wetting layer on 
  each single sphere which is a remnant of the prewetting transition
  on the corresponding flat substrate. Thus one has to distinguish
  between the prewetting line for a first-order wetting transition on
  a planar substrate, the thin-thick transition line for wetting on a
  single sphere, and the bridging transition line for two spheres
  (Fig.~\ref{f:btandttt}). At two-phase coexistence $\Delta\mu=0$ and
  for $R\gg\sigma$ the bridging transition is determined by the equality of the
  surface areas of the interfaces in the bridged and the unbridged configuration,
  leading to a universal ratio $D_{bt}(\Delta\mu=0)/(R+l_0)\approx2.32$ for the
  critical distance $D_{bt}(\Delta\mu=0)$ of the bridging transition
  at coexistence (Fig.~\ref{f:eip_largespheres} and
  Subsec.~\ref{ss:eipforlargespheres}). Off 
  coexistence $D_{bt}(\Delta\mu,R)$ is described by a universal
  scaling function (Eq.~(\ref{e:Lambda})).
\item At large distances and depending on the temperature relative to
  the thin-thick transition temperature on a single sphere the
  wetting-induced effective interaction potential can be either
  attractive or repulsive; in both cases it decays $\sim D^{-6}$ for
  large $D$. The bridging transition leads to a strong break in slope of the
  effective interaction potential at $D=D_{bt}$. This is the
  fingerprint of a cooperative phenomenon among the fluid particles
  whose degrees of freedom have been integrated out (see
  Sec.~\ref{s:intro}). Metastable branches
  of the effective potential give rise to pronounced hysteresis
  effects (Fig.~\ref{f:eipforexample1}).
\item In the case that a bridge of the wetting phase connects the
  spheres (i.e., $D<D_{bt}$) there is an attractive wetting-induced interaction
  (Fig.~\ref{f:eipforexample1}) that pulls the
  spheres together. Within a wide range of separations $a=D-2R$ of the
  spherical surfaces this force is of the same order of magnitude as
  the bare dispersion interaction potential $\Phi$
  (Eq.~(\ref{e:vdwbetwsph})) between the spheres. This bare
  interaction of two spheres (corresponding to the case that they are
  separated by vacuum) has to be added to
  the effective potential $\Omega_S$ to yield the total interaction
  potential $\Omega_{tot}$ between the spheres which is attractive at
  large distances
  (Eqs.~(\ref{e:vdwbetwsph}), (\ref{e:omegatot}), and
  Fig.~\ref{f:eipforexample1plusbare}).  
\item The wetting-induced force between spherical particles is
  experimentally accessible \emph{directly} through suitable force
  microscopy techniques (Subsec.~\ref{ss:forcemicroscopy}). Moreover,
  in Subsec.~\ref{ss:sscs} we argue 
  that this force influences the phase behavior of \emph{stericly}
  stabilized, index-matched colloidal suspensions. The total effective
  interaction potential for such a case is shown in
  Fig.~\ref{f:indexmatch}; it is repulsive at large
  distances. The phase behavior of \emph{charge} stabilized 
  colloidal suspensions (Subsec.~\ref{ss:cscs}) is only affected by
  the wetting-induced interaction potential if the screening length of the Coulomb
  repulsion in the solvent is smaller than
  $a_{bt}=D_{bt}-2R\approx0.32R$. Depending on the size of the
  charges, the salt concentration of the solvent, and the underlying
  dispersion forces this criterion may be fulfilled even for colloidal
  particles whose radii are only a few tens of nm.
\end{enumerate}

\acknowledgements

We gratefully acknowledge financial support by the German Science
Foundation within the special research initiative
\emph{Wetting and Structure Formation at Interfaces}.

\appendix

\section*{Contributions to the free energy}

Our choice of interaction potentials $\phi(r)$ (Eq.~(\ref{e:ljff}))
and $\phi_s(r)$ (Eq.~(\ref{e:ljsphere})) leads to the following
expressions for the contributions to the free energy
$\Omega_S$ (with the thermodynamic limit already carried out):

\begin{equation}
\Omega_{ei}[h(z)] = 2\Delta\rho\left(\rho_l \int\limits_0^{\infty}
  dz\, (g_+(z)+g_-(z)) - \int\limits_0^{\infty} dz\,
  (f_+(z)+f_-(z))\right)
\end{equation}
with
\begin{eqnarray}
g_{\pm}(z) & = & 2w_0\sigma^2\frac{R_1}{\sigma} -
    w_0\sigma^2\left(\frac{h^2(z)}{\sigma^2} +
    \left(\frac{z}{\sigma}\pm\frac{D}{2\sigma}\right)^2 -
    \frac{R_1^2}{\sigma^2}+1\right) \\ 
& &
    \times\left(\arctan\left(\sqrt{\frac{h^2(z)}{\sigma^2} +
    \left(\frac{z}{\sigma}\pm\frac{D}{2\sigma}\right)^2}
    + \frac{R_1}{\sigma}\right)\right. - \nonumber \\
& & 
    \left.\arctan\left(\sqrt{\frac{h^2(z)}{\sigma^2}+\left(\frac{z}{\sigma}
    \pm\frac{D}{2\sigma}\right)^2} 
    - \frac{R_1}{\sigma}\right)\right) \nonumber
\end{eqnarray}
where $R_1 = R+d_s$ and
\begin{eqnarray}
f_{\pm}(z) & = & \frac{\pi u_9}{4}\left(\frac{1}{7}\left
    ( \frac{1}{(k_{\pm}+R)^7} - \frac{1}{(k_{\pm}-R)^7}\right) + R
    \left(\frac{1}{(k_{\pm}+R)^8} +
    \frac{1}{(k_{\pm}-R)^8}\right)\right) \nonumber\\
& & - \pi u_3 \left( \frac{1}{k_{\pm}+R} - \frac{1}{k_{\pm}-R} + R
    \left(\frac{1}{(k_{\pm}+R)^2} +
    \frac{1}{(k_{\pm}-R)^2}\right)\right)
\end{eqnarray}
with $k_{\pm} = \sqrt{h^2(z)+(z\pm D/2)^2}$. The contribution
$\Omega_{lg}$ is given by
\begin{eqnarray}\label{e:app_omegalg}
\Omega_{lg}[h(z)] & = & -w_0\sigma^3(\Delta\rho)^2 \int\limits_0^{\infty} dz
\int\limits_0^{\infty} dz' \,\left(
\frac{1}{y_-^4}\left(\frac{qy_-^6+y_-^4(2q^2+p^2)+3p^2qy_-^2+p^4}
  {(p^2+2y_-^2q+y_-^4)^{3/2}} - p\right)\right. \nonumber\\
& & + \left.\frac{1}{y_+^4}\left(\frac{qy_+^6+y_+^4(2q^2+p^2)+3p^2qy_+^2+p^4}
  {(p^2+2y_+^2q+y_+^4)^{3/2}} - p\right)\right)
\end{eqnarray}
where the abbreviations $y_{\pm}$, $p$, and $q$ are defined by
\begin{equation}
y_{\pm}^2 = \sigma^2+(z\pm z')^2,
\end{equation}
\begin{equation}
p = h^2(z)-h^2(z'),
\end{equation}
and
\begin{equation}
q = h^2(z)+h^2(z').
\end{equation}
The double integral in Eq.~(\ref{e:app_omegalg}) demonstrates the
nonlocal functional dependence of $\Omega_{lg}$ on $h(z)$.

\newpage

\newpage

\end{document}